# Towards Shift Tolerant Visual Secret Sharing Schemes

Daoshun Wang[a,*], Lin Dong[a], Xiaobo Li[b]

[a]Department of Computer Science and Technology, Tsinghua University, Beijing, 100084, China

[b]Department of Computing Science, University of Alberta, Edmonton, Alberta, T6G 2E8, Canada

**Abstract**

In (*k*, *n*) visual secret sharing (VSS) scheme, secret image can be visually reconstructed when *k* or more participants printing theirs shares on transparencies and stack them together. No secret is revealed with fewer than *k* shares. The alignment of the transparencies is important to the visual quality of the reconstructed secret image. In VSS scheme, each pixel of the original secret image is expanded to *m* sub-pixels in a share image. If a share image is printed on paper with the same size as the original secret image, the alignment or the registration of the sub-pixels, which is only *m* times smaller than that in the original secret, could be troublesome. Liu et al. [4] has noticed this alignment problem and observed that some information of the secret image may still be revealed even when the shares are not precisely registered in the horizontal direction. Yang et al. [9] introduced a general approach to construct a misalignment tolerant (*k*, *n*)-VSS scheme using big and small blocks for the situation where the original secret image has a certain degree of redundancy in shape accuracy. In this paper, we propose a (2, *n*)-VSS scheme that allows a relative shift between the shares in the horizontal direction and vertical direction. When the shares are perfectly aligned, the contrast of the reconstructed image is equal to that of traditional VSS. When there is a shift, average contrast in the reconstructed image is higher than that of the traditional VSS, and the scheme can still work in the cases where very little shape redundancy presents in the image. The trade-off is that our method involves a larger pixel expansion. The basic building block of our scheme is duplication and concatenation of certain rows or columns of the basic matrices. This seemingly simple but very powerful construction principle can be easily used to create more general (*k*, *n*)-VSS schemes.

***Keywords*****:** Visual secret sharing, pixel expansion, share alignment and registration.

## 1. Introduction

* Corresponding author. Tel.:+86-10-62785636.
*E-mail address:* daoshun@mail.tsinghua.edu.cn (Daoshun Wang).

A visual secret sharing (VSS) scheme (or visual cryptography) has been proposed to encode a secret image $S$ into $n$ “shadow” images (or “shares”) to be distributed among $n$ participants [1]. In a ($k$, $n$)-VSS scheme, each pixel of the secret image, which is unit element, is “expanded” into $m$ sub-pixels in each share. The secret image $S$ can be visually reconstructed with $k$ shares, printed on transparencies, superimposed precisely on an overhead projector. A slight misalignment between the shares could dramatically degrade the visual quality of the reconstructed image. A great effort has been directed to reduce the size of the share transparencies (e.g., by reducing $m$), but smaller transparencies make the alignment more difficult, even with the help of printed “frame” or “makers”. This alignment or registration problem is made worse if some distortion is involved in the process of producing transparencies, either by photocopying or laser printing. The heat produced in the printing process may bend the plastic sheets.

Kobara and Imai [2] considered the visibility problem of the decoded image in (2, 2)-VSS scheme when the viewpoint is changed, and categorized the space where the viewpoint belongs according to the visibility. They gave a relation between the shift of two corresponding cells and its density (visibility) in (2, 2)-VSS Scheme. Nakajima and Yamaguchi [3] proposed a (2, 2)-extended VSS scheme which can enhance registration tolerance when stacked shares do not align perfectly.

In practice, stacking more than two or three transparencies accurately is quite difficult. In a general $(k, n)$-VSS scheme, the pixel expansion is $m \in \left[ \binom{n-1}{k-1} 2^{k-2} + 1, \binom{n-1}{k-1} 2^{k-1} + 1 \right]$, as pointed out by Blundo et al. [6]. Since the optimal contrast for a $(k, n)$-scheme is $1/m$, this contrast is quite small and the visual reconstruction of the secret image can be very hard. Therefore, one only recovers secret image by resorting to computers when $k$ is greater than three. For this reason, most visual schemes mainly handle the cases with $k$ being at most 3. Here,we only focus on the case of $k = 2$ because the contrast of a (2, $n$)-VSS scheme is much better than that of a $(k, n)$-VSS scheme with $k \geq 3$. (2, $n$)-VSS scheme is significantly more practical in real-life visual reconstruction.

Liu et al. [4] observed that the secret image can still be revealed visually even when the transparencies are not aligned perfectly in the horizontal direction. They analyzed the alignment problem of a binary (2, 2)-VSS scheme and obtained a nice relation between the horizontal shift, $x$, of the shares and the average contrast of the reconstructed secret image $\overline{a}$ as $\overline{a} = -(m - x)e/(m^2(m-1))$, where $e = h - l$ ([5]). In reality, the misalignment happens not only horizontally, thus the shifts in the vertical and diagonal directions also need to be addressed. Yang et al. [9] introduced a general approach to construct a misalignment tolerant ($k$, $n$)-VSS scheme using big and small blocks for the situation where the original secret image has a certain degree of redundancy in shape accuracy. The secret image can still be recovered under slight misalignment, and the quality of the reconstructed image is assessed by manual inspection. The method proposed in Ref. [9] has certain difficulties when the secret image only contains thin lines. The significant advantage of Yang's scheme is that it has no pixel expansion.

Inspired by the works of Ref. [4] and Ref. [9], we propose, in this paper, a shift tolerant VSS that provides higher quality reconstructed images than the traditional schemes, and it works better than scheme in Ref. [9] for the cases where very little shape redundancy exists in the secret image. When the shares are perfectly aligned, the secret image is precisely recovered. When there is a shift, the image quality can be measured by the average contrast. The construction principle is also used to create more general $(k, n)$-VSS schemes.

The rest of this paper is organized as follows. Section 2 reviews the alignment problem of VSS scheme. Section 3 first gives a formal definition of shift tolerant VSS scheme and then proposes a shift tolerant $(2, n)$-VSS scheme. A general approach to construct shift tolerant $(k, n)$-VSS scheme is given in Section 4. Comparisons and discussions are given in Section 5, and the conclusions are given in Section 6.

## 2. Background, preliminaries and motivation

This section briefly reviews the traditional VSS schemes ([1, 5]) and the share alignment issue ([4]). And motivation to deal with this alignment issue is introduced. Some basic notations are defined when they first appear in the text and a list of important notations is given in Appendix A.

### *2.1 Binary (k, n)-VSS scheme schemes*

In binary VSS scheme, the secret image consists of a collection of black–and-white pixels and each pixel is subdivided into a collection of *m* black–and-white sub-pixels in each of the *n* shares. The collection of sub-pixels can be represented by an $n \times m$ Boolean matrix $S = [s_{ij}]$, where the element $s_{ij}$ represents the *j-th* sub-pixel in the *i-th* share. A white pixel is represented as a 0, and a black pixel is represented as a 1. On a transparency, white sub-pixels allow light to pass through while black sub-pixels stop light. $s_{ij} = 1$ if and only if the *j-th* pixel in the *i-th* share is black. Stacking shares $i_1, \ldots, i_r$ together, the grey-level of each pixel (*m* sub-pixels) of the combined share is proportional to the Hamming weight (the number of 1's in the vector *V*) $H(V)$ of the OR-ed ("OR" operation) *m*-vector $V = OR(i_1, \mathbf{L}, i_r)$ where $i_1, \ldots, i_r$ are the rows of *S* associated with the shares we stack. Verheul and Van Tilborg[5] extended the definition of Naor and Shamir's scheme[1].

The formal definition of binary VSS scheme is given below.

**Definition 1.** [5] A solution to the $k$ out of $n$ visual secret sharing scheme consists of two collections of $n \times m$ Boolean matrices $C_0$ and $C_1$. To share a white (resp. black) pixel, the dealer randomly chooses one of the matrices in $C_0$ (resp. $C_1$). The chosen matrix defines the color of the $m$ sub-pixels in each one of the $n$ transparencies. The solution is considered valid if the following three conditions are met.

1. For any $S$ in $C_0$, the OR vector $V^0$ of any $k$ of the $n$ rows satisfies $H(V^0) \le l, l \in Z^+$.
2. For any $S$ in $C_1$, the OR vector $V^1$ of any $k$ of the $n$ rows satisfies $H(V^1) \ge h, h \in Z^+, l < h \le m$.
3. For any subset $\{i_1, \ldots, i_q\}$ of $\{1, \ldots, n\}$ with $q < k$, the two collections of $q \times m$ matrices $D_t$ for $t \in \{0, 1\}$ obtained by restricting each $n \times m$ matrix in $C_t$ (where $t = 0, 1$) to rows $i_1, \ldots, i_q$ are indistinguishable in the sense that they contain the same matrices with the same frequencies.

The first two conditions are called "contrast" and the third condition is called "security". In this definition the parameter $m$ is called "pixel expansion", which refers to the number of sub-pixels representing a pixel in the secret image. The contrast $a = \frac{H(V^1) - H(V^0)}{m} = \frac{h - l}{m}$, also called relative contrast difference, refers to the difference in weight between combined shares that come from a white pixel and a black pixel in the secret image.

From Definition 1, a binary ($k$, $n$)-VSS scheme can be realized by the two Boolean matrices $B_0$ and $B_1$. The collection $C_0$ (resp. $C_1$) can be obtained by permuting the columns of the corresponding Boolean matrix $B_0$ (resp. $B_1$) in all possible ways. $B_0$ and $B_1$ are called basis matrices, and hence each collection has $m!$ matrices.

An example (2, 3) scheme is given below as a demonstration for Definition 1.

**Example 1.** Suppose in (2, 3)-VSS scheme, each pixel in a secret image is encoded into a collection of 3 black and white sub-pixels in each of the 3 shares. The encoding matrices can be represented by two $3 \times 3$ 0/1 matrices (see Ref. [8] ):

$$B_0 = \begin{bmatrix} 0 & 1 & 1 \\ 0 & 1 & 1 \\ 0 & 1 & 1 \end{bmatrix} \qquad B_1 = \begin{bmatrix} 0 & 1 & 1 \\ 1 & 0 & 1 \\ 1 & 1 & 0 \end{bmatrix}.$$

For a white (resp. black) pixel in a secret image, we encode it as three sub-pixels for each of the three shares, corresponding to the first, second and third row of $B_0$ (resp. $B_1$ ), respectively.

Hamming weight of any one row in $B_0$ and $B_1$ is 2, every secret pixel in each share is represented as two black pixel and one white pixel and security condition is satisfied.

The Hamming weight of the OR $m$-vector $V^0(V^1)$ of the first row and second row in $B_0$ ($B_1$) is $H(V^0)=2$ ($H(V^1)=3$). The pixel expansion $m$=3, and contrast $a=(H(V^1)-H(V^0))/m=1/3$.

### *2.2 The alignment problem of VSS scheme*

Liu et al. [4] found the phenomenon that the precise alignment of small pixels is not critical. A recognizable secret image (with some quality loss) can still be recovered visually even if the participants do not align the transparencies precisely in VSS scheme.

Their shifted scheme is generated as follows: shift the second row of the $m!$ share matrices in $C_0$ (resp. $C_1$) to the left (resp. right) by $x$ pixels, and then $m!$ shifted share matrices are gotten.

**Theorem 1[4]:** In a conventional (2, 2)-VSS scheme with respect to pixel expansion $m$, when share 2 and share 1 are stacked together precisely, the scheme has contrast $a=\frac{h-l}{m}$.When the share 2 is shifted $x\,(1\le x\le (m-1))$ pixels position relative to share 1 in horizontal direction, the average contrast $\overline{a}=-(m-x)e/(m^2(m-1))$, here $e=h-l$.

We use the typical traditional (2, 2)-VSS scheme to demonstrate its behavior in both situations of perfect alignment and share shifts.

**Example 2.** The basis matrices of the (2, 2)-VSS scheme are as follows:

$$B_0=\begin{bmatrix}1 & 0\\ 1 & 0\end{bmatrix},\quad B_1=\begin{bmatrix}1 & 0\\ 0 & 1\end{bmatrix}.$$

When the two shares are stacked together precisely, the contract is $a=1/2$.

When share 2 (denote by $S_2$) is shifted to left (or right) by 1 pixel relative to share 1(denote by $S_1$), all possible cases are showed in Table 1. The pixel shifted in could be either 0 or 1, which is denoted by *c*. The pixel shifted out is denoted by "*". Let $p_0$ be the probability that a 0 is shifted in, and $p_1$ be the probability that a 1 is shifted in, we have $p_0=p_1=0.5$. $\overline{l}$ (resp. $\overline{h}$) represents average value of $l$ (resp. $h$). The average contrast is $\overline{a}=-1/4$. It is easily to show that the average contrast $\overline{a}$ in Table 1 is the same by using Theorem 1.

**Table 1.** All possible situations when $S_2$ is shifted to the left by one pixel relative to $S_1$

| Pixel | Prob. | Matrices | Shifted matrices | Average Hamming weight of the stacked vector | Average contrast |
|---|---|---|---|---|---|
| | 0.5 | $\begin{bmatrix}1&0\\1&0\end{bmatrix}$ | $\begin{bmatrix}*&1&0\\1&0&\boldsymbol{c}\end{bmatrix}$ | $\bar{l}=0.5\times(1\times p_0+2\times p_1)+0.5\times(2\times p_0+2\times p_1)=1.75$ | $\bar{a}=\frac{\bar{h}-\bar{l}}{m}=\frac{1.25-1.75}{2}=-\frac{1}{4}$ |
| | 0.5 | $\begin{bmatrix}0&1\\0&1\end{bmatrix}$ | $\begin{bmatrix}*&0&1\\0&1&\boldsymbol{c}\end{bmatrix}$ | | |
| | 0.5 | $\begin{bmatrix}1&0\\0&1\end{bmatrix}$ | $\begin{bmatrix}*&1&0\\0&1&\boldsymbol{c}\end{bmatrix}$ | $\bar{h}=0.5\times(1\times p_0+2\times p_1)+0.5\times(1\times p_0+1\times p_1)=1.25$ | |
| | 0.5 | $\begin{bmatrix}0&1\\1&0\end{bmatrix}$ | $\begin{bmatrix}*&0&1\\1&0&\boldsymbol{c}\end{bmatrix}$ | | |

When share 2 is shifted up (or down) by 1 pixel relative to share 1, the original row 2 is shifted out and a new row is shifted in. The pixels shifted in maybe 01 or 10. All possible cases are shown in Table 2. The average contrast is $\bar{a}=0$.

**Table 2.** All possible situations when $S_2$ is shifted up by one pixel relative to $S_1$

| Pixel | Prob. | Matrices | Hamming weight of Shifted matrices | Average Hamming weight of the stacked vector | Average contrast |
|---|---|---|---|---|---|
| | 0.5 | $\begin{bmatrix}1&0\\1&0\end{bmatrix}$ | $H([10]+[10])=1$ or $H([10]+[01])=2$ | $\bar{l}=0.5\times(0.5\times1+0.5\times2)+0.5\times(0.5\times1+0.5\times2)=1.5$ | $\bar{a}=\frac{\bar{h}-\bar{l}}{m}=\frac{1.5-1.5}{2}=0$ |
| | 0.5 | $\begin{bmatrix}0&1\\0&1\end{bmatrix}$ | $H([01]+[10])=1$ or $H([01]+[01])=2$ | | |
| | 0.5 | $\begin{bmatrix}1&0\\0&1\end{bmatrix}$ | $H([10]+[10])=1$ or $H([10]+[01])=2$ | $\bar{h}=0.5\times(0.5\times1+0.5\times2)+0.5\times(0.5\times1+0.5\times2)=1.5$ | |
| | 0.5 | $\begin{bmatrix}0&1\\1&0\end{bmatrix}$ | $H([01]+[10])=1$ or $H([01]+[01])=2$ | | |

It is clear that in a conventional (2, 2)-VSS scheme, one cannot obtain any information of the secret image when the shares are shifted vertically by one pixel, relative to the other share. One example secret image and its reconstruction results are shown in Fig.1.

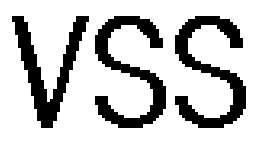

(a) Original secret image

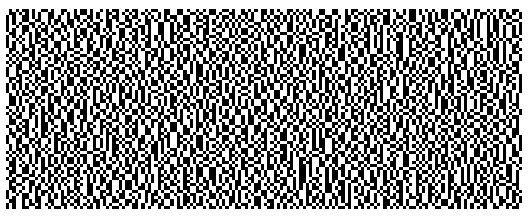

(b) $S_1$

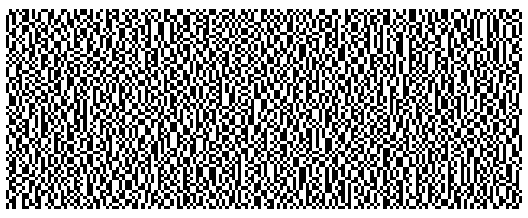

(c) $S_2$

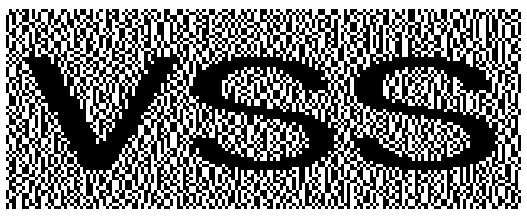

(d) $S_1$ and $S_2$ are stacked precisely

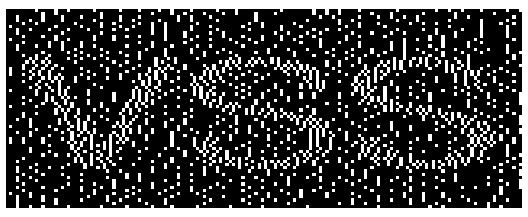

(e) $S_2$ is shifted left by 1 pixel

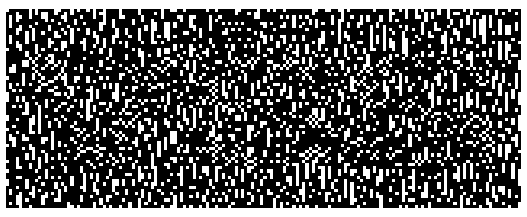

(f) $S_2$ is shifted up by 1 pixel

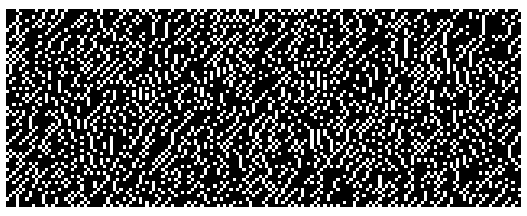

(g) $S_2$ is shifted left and down by 1 pixel

**Fig. 1** The stacking results of the (2, 2)-VSS scheme

We can see that when $S_1$ and $S_2$ are stacked precisely, the reconstructed secret image has a nice contrast. When $S_2$ is shifted left by one pixel relative to $S_1$, secret image can be reconstructed by human visual system; however, the recovered secret image is the complementary image of the original one. When $S_2$ is shifted up by one pixel relative to $S_1$, the secret image cannot be recognized by human visual system at all.

The secret image of the conventional VSS scheme can still be recovered visually when a small amount of horizontal displacement exists. The absolute value of $\bar{a}$ becomes smaller, and it is harder for human being to recognize the recovered secret image visually. If the shared shares do not align perfectly vertically or diagonally, visual reconstruction is impossible. In this situation, our aim to design shift tolerant VSS scheme, the scheme has two properties, one property is that it is fully compatible to traditional VSS scheme with respect to quality of reconstructed secret image, and other is visual reconstruction of secret image when the stacked shares are slight misalignment.

### *2.3 Motivation*

The simplest method for shift-tolerance is to make the share images bigger, which is equivalent to simply duplicating the sub-pixels. Using the scheme in Example 2, we trivially obtain the following scheme, in Example 3, by duplicating pixels, which results in a lager pixel expansion: $m$ =4.

**Example 3**. The basis matrices of the (2, 2)-VSS scheme with "pixel duplication" are as follows:

$$B_0 = \begin{bmatrix} 1 & 1 & 0 & 0 \\ 1 & 1 & 0 & 0 \end{bmatrix}, \quad B_1 = \begin{bmatrix} 1 & 1 & 0 & 0 \\ 0 & 0 & 1 & 1 \end{bmatrix}.$$

Table 3 gives the calculation for the average contrast in the case of horizontal shift of one pixel.

**Table 3.** Duplicating pixels of basis matrices horizontally with horizontal shift of one pixel

| Pixel | Prob. | Matrices | Shifted matrices | Average Hamming weight of the stacked vector | Average contrast |
|---|---|---|---|---|---|
| | 0.5 | $\begin{bmatrix}1100\\1100\end{bmatrix}$ | $\begin{bmatrix}* & 1100\\1 & 100c\end{bmatrix}$ | $\bar{l} = 0.5\times(2\times p_0 + 3\times p_1) + 0.5\times(3\times p_0 + 3\times p_1) = 2.75$ | $\bar{a} = \frac{\bar{h}-\bar{l}}{m} = \frac{3.25-2.75}{4} = \frac{1}{8}$ |
| | 0.5 | $\begin{bmatrix}0011\\0011\end{bmatrix}$ | $\begin{bmatrix}* & 0011\\0 & 011c\end{bmatrix}$ | | |
| | 0.5 | $\begin{bmatrix}1100\\0011\end{bmatrix}$ | $\begin{bmatrix}* & 1100\\0 & 011c\end{bmatrix}$ | $\bar{h} = 0.5\times(3\times p_0 + 4\times p_1) + 0.5\times(3\times p_0 + 3\times p_1) = 3.25$ | |
| | 0.5 | $\begin{bmatrix}0011\\1100\end{bmatrix}$ | $\begin{bmatrix}* & 0011\\1 & 100c\end{bmatrix}$ | | |

It can be seen that with the shares twice as big, the result average contrast of the reconstructed image is even lower than the original scheme, as far the absolute values of the contrasts are concerned. In other words, simply enlarge the shares may not solve the shift problem.

We now consider a different duplication mechanism, in Example 4. Instead of duplicating pixels, we duplicate the vectors of the original basis matrices.

**Example 4**. The basis matrices of the (2, 2)-VSS scheme with "row vector duplication" are as follows:

$$B_0 = \begin{bmatrix} 1 & 0 & 1 & 0 \\ 1 & 0 & 1 & 0 \end{bmatrix}, \quad B_1 = \begin{bmatrix} 0 & 1 & 0 & 1 \\ 1 & 0 & 1 & 0 \end{bmatrix}.$$

Table 4 gives the calculation of the average contrast in the case of horizontal shift of one pixel.

**Table 4.** Duplicating vectors of basis matrices horizontally with horizontal shift of one pixel

| Pixel | Prob. | Matrices | Shifted matrices | Average Hamming weight of the stacked vector | Average contrast |
|---|---|---|---|---|---|
| | 0.5 | $\begin{bmatrix} 1010 \\ 1010 \end{bmatrix}$ | $\begin{bmatrix} * & 1010 \\ 1 & 010c \end{bmatrix}$ | $\bar{l} = 0.5\times(3\times p_0 + 4\times p_1) + 0.5\times(4\times p_0 + 4\times p_1) = 3.75$ | $\bar{a} = \frac{\bar{h} - \bar{l}}{m} = \frac{2.25 - 3.75}{4} = -\frac{3}{8}$ |
| | 0.5 | $\begin{bmatrix} 0101 \\ 0101 \end{bmatrix}$ | $\begin{bmatrix} * & 0101 \\ 0 & 101c \end{bmatrix}$ | | |
| | 0.5 | $\begin{bmatrix} 1010 \\ 0101 \end{bmatrix}$ | $\begin{bmatrix} * & 1010 \\ 0 & 101c \end{bmatrix}$ | $\bar{h} = 0.5\times(2\times p_0 + 3\times p_1) + 0.5\times(2\times p_0 + 2\times p_1) = 2.25$ | |
| | 0.5 | $\begin{bmatrix} 0101 \\ 1010 \end{bmatrix}$ | $\begin{bmatrix} * & 0101 \\ 1 & 010c \end{bmatrix}$ | | |

It is cleat that duplicating vectors results in an increase in absolute value for the average contrast. Table 5 uses one example secret image to demonstrate the advantage of duplicating vectors in the above examples. More detailed analysis of Example 3 and Example 4 under different shift situations is given in Appendix B.

**Table 5**. Comparison of "pixel duplication" and "row vector duplication"

| | Pixel duplication | Row vector duplication |
|---|---|---|
| Stacking | $B_0 = \begin{bmatrix} 1 & 0 \\ 1 & 0 \end{bmatrix} \Rightarrow B_0 = \begin{bmatrix} 1 & 1 & 0 & 0 \\ 1 & 1 & 0 & 0 \\ 1 & 1 & 0 & 0 \\ 1 & 1 & 0 & 0 \end{bmatrix}$ $B_1 = \begin{bmatrix} 1 & 0 \\ 0 & 1 \end{bmatrix} \Rightarrow B_1 = \begin{bmatrix} 1 & 1 & 0 & 0 \\ 1 & 1 & 0 & 0 \\ 0 & 0 & 1 & 1 \\ 0 & 0 & 1 & 1 \end{bmatrix}$ | $B_0 = \begin{bmatrix} 1 & 0 \\ 1 & 0 \end{bmatrix} \Rightarrow B_0 = \begin{bmatrix} 1 & 0 & 1 & 0 \\ 1 & 0 & 1 & 0 \\ 1 & 0 & 1 & 0 \\ 1 & 0 & 1 & 0 \end{bmatrix}$ $B_1 = \begin{bmatrix} 1 & 0 \\ 0 & 1 \end{bmatrix} \Rightarrow B_1 = \begin{bmatrix} 1 & 0 & 1 & 0 \\ 1 & 0 & 1 & 0 \\ 0 & 1 & 0 & 1 \\ 0 & 1 & 0 & 1 \end{bmatrix}$ |
| Precise alignment | 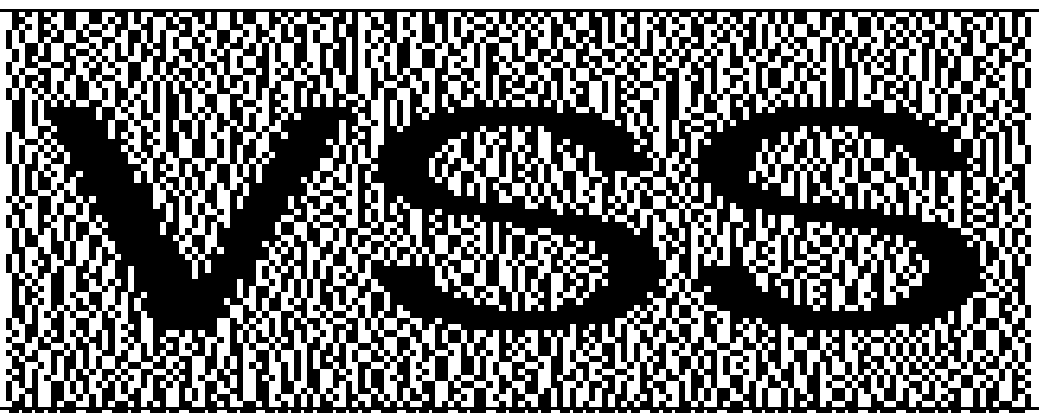 | 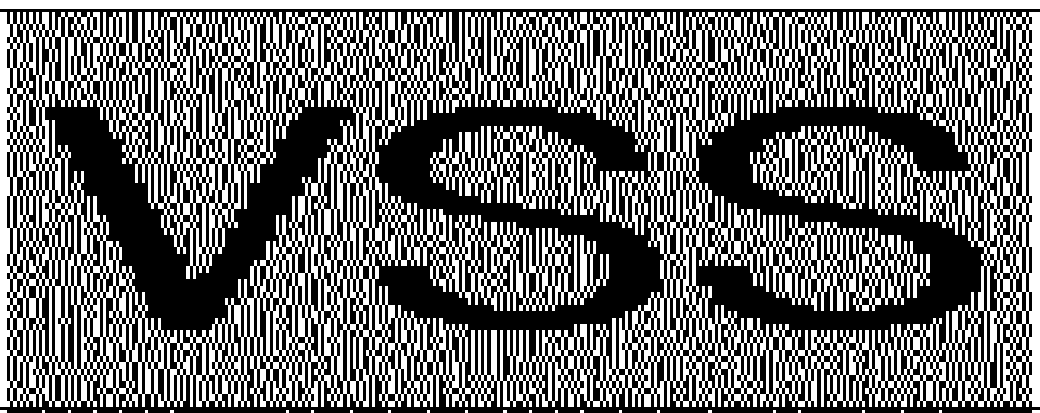 |
| $S_2$ is shifted left by 1 pixel | 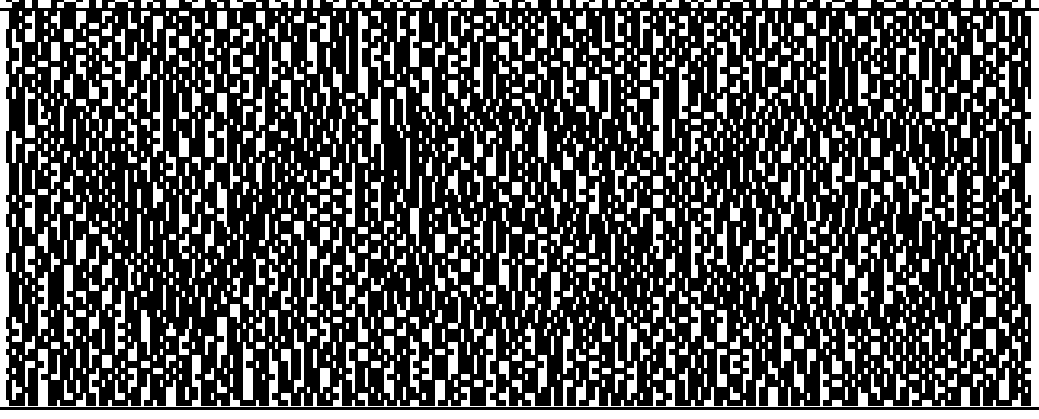 | 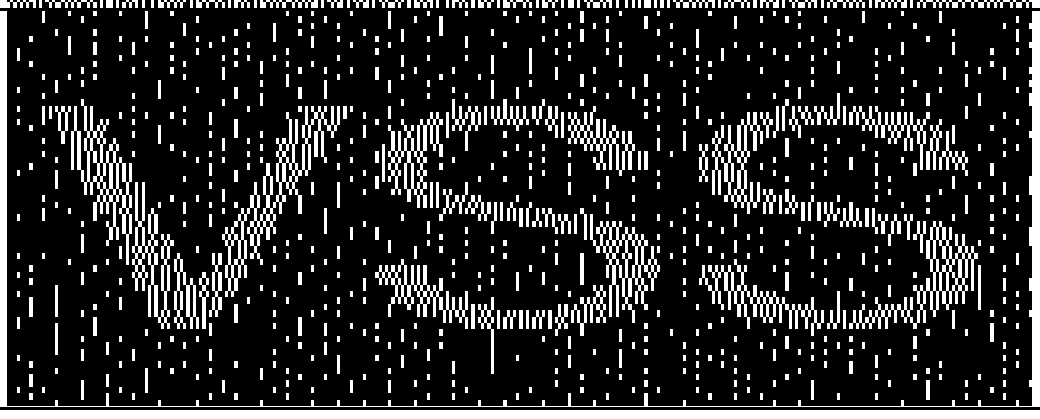 |

Building on top of this vector duplication concept, we propose, in the next section, a new VSS scheme that can provide better recovered image in the case of share misalignment.

# 3. Shift Tolerant Visual Secret Sharing Scheme

In this section, we propose new approach to construct visual secret sharing scheme (donated by STVSS scheme) that allows minor displacement in horizontal and vertical directions between the shares.

## *3.1 Definition of shift tolerant visual secret sharing scheme*

In basis matrix of the conventional VSS scheme, each row vector with $m$ sub-pixels corresponds to one share. In our scheme, $N_y (N_y \in Z^+)$ row vectors of the basis matrix are corresponds to one share. And each row vector is composed of $N_x (N_x \in Z^+)$ row vectors of the basis matrix of the conventional VSS scheme.

Each pixel of the original secret image is represented by $N_y N_x m$ black and white sub-pixels in each of the $n$ shares. The collection of sub-pixels is represented by a $N_y n \times N_x m$ Boolean matrix $S' = [\, s'_{i'j'} \,]$. where the element $s'_{i'j'}$ represents the $j'$-th sub-pixel in the $i'$-th share, and $i' = N_y \cdot i$, $j' = N_x \cdot j, i \in [1, n], j = [1, m]$. We have $s'_{i'j'}$ =0 if and only if the $j'$-th sub-pixel in the $i'$-th share is white, while $s'_{i'j'}$ = 1 if and only if that sub-pixel is black.

The "OR" operation of the pixels within row vectors models the stacking operation of sub-pixels, just the same as in a conventional VSS scheme.

Let $V^t$ be a group /block of $N_y$ row vectors (where $t$ = 0, 1),

$$V^t = \begin{bmatrix} V^t[1] \\ \mathbf{L} \\ V^t[N_y] \end{bmatrix}.$$

Each $V^t[i]$ is an $m_i$-vector and the number of 1's in $V^t[i]$ is called the Hamming weight of $V^t[i]$, denoted by $H(V^t[i])$, $i = 1, \mathbf{L}, N_y$. Hamming weight of $V^t$ can be computed in this way:

$$H(V^t) = H(\begin{bmatrix} V^t[1] \\ \mathbf{L} \\ V^t[N_y] \end{bmatrix}) = \begin{bmatrix} H(V^t[1]) \\ \mathbf{L} \\ H(V^t[N_y]) \end{bmatrix}.$$

When $N_y = 1$, $H(V^t)$ is a scalar and it reduces to the Hamming weight defined in Ref. [1].

In the formal definition of our STVSS scheme below, we use symbol $x$ to denote the deviation in horizontal, and use $y$ to denote the deviation in vertical.

We now give a formal definition of the proposed STVSS scheme.

**Definition 2**. A $k$ out of $n$ shift-tolerant visual secret sharing (STVSS) scheme consists of two collections of $N_y n \times N_x m$ Boolean matrices $C_0^*$ and $C_1^*$. To share a white (black) pixel, the dealer randomly chooses one of the matrices in $C_0^*$ ($C_1^*$). The chosen matrix defines the color of the $m^*$ ($N_y N_x m$) sub-pixels in each one of the $n$ shares (transparencies). This scheme is considered valid if the following conditions are met.

1. The "Contrast" condition when $x, y = 0$:

   This is the case that any $k$ of $n$ shares are perfectly aligned to each other, no displacements. For any S in $C_0^*$, the OR vector $V^0$ of any $k$ of the $n$ blocks of row vectors satisfy $H(V^0) \le N_y N_x l$. For any S in $C_1^*$, the OR vector $V^1$ of any $k$ of the $n$ blocks of row vectors satisfy $H(V^1) \ge N_y N_x h$.

2. The "(Average) Contrast" condition when $1 \le x \le m$ and $1 \le y \le N_y$:

   This is the case that one share is shifted $x$ pixels horizontally and $y$ pixel vertically relative to the other $k-1$ shares. For any S in $C_0^*$ shifted, the OR vector $V^0$ of any $k$ of the $n$ blocks of row vectors satisfy $\overline{H(V^0)} \le \bar{l}$. For any S in $C_1^*$ shifted, the OR vector $V^1$ of any $k$ of the $n$ blocks of row vectors satisfy $\overline{H(V^1)} \ge \bar{h}$.

3. The "Security" condition:

   For any subsets $\{i_1, \ldots, i_q\}$ of $\{1, \ldots, n\}$ with $q < k$, the two collections of $qN_y \times m^*$ matrices $D_t^*$ for $t \in \{0, 1\}$ obtained by restricting each $n^* \times m^*$ matrix in $C_t^*$ (where $t$ = 0, 1) to a group of rows $i_1, \ldots, i_q$ are indistinguishable in the sense that they contain the same matrices with the same frequencies.

In Definition 2 above, the parameter $m^*$ is called the pixel expansion, which refers to the number of sub-pixels representing one pixel in the original secret image. The collections of Boolean matrices $C_0^*$ and $C_1^*$ can be obtained by permuting the columns of matrices $B_0^*$ and $B_1^*$, which are called basis matrices.)

We obtain $a^{(x,y)}$ as:

$$a^{(x,y)} \cdot m^* = H(V^1) - H(V^0) = H(\begin{bmatrix} V^1[1] \\ \mathbf{L} \\ V^1[N_y] \end{bmatrix}) - H(\begin{bmatrix} V^0[1] \\ \mathbf{L} \\ V^0[N_y] \end{bmatrix})$$

$$=\begin{bmatrix} H(V^1[1]) & -H(V^0[1]) \\ \mathbf{L} & \mathbf{L} \\ H(V^1[N_y]) - & H(V^0[N_y]) \end{bmatrix} = \begin{bmatrix} a_1 \cdot m_1 \\ \mathbf{L} \\ a_{N_y} \cdot m_{N_y} \end{bmatrix}.$$

By definition of Hamming weight, the value of $\begin{bmatrix} a_1 \cdot m_1 \\ \mathbf{L} \\ a_{N_y} \cdot m_{N_y} \end{bmatrix}$ equals to value of $a_1 \cdot m_1 + \mathbf{L} + a_{N_y} \cdot m_{N_y}$.

In conventional VSS scheme, $a_1 = a_2 = \mathbf{L} = a_{N_y} = \frac{h-l}{m}$. We have $m_1 = m_2 = \mathbf{L}\, m_{N_y} = N_x m$. When $x=0$ and $y=0$, $a^{(0,0)} = \frac{N_y N_x (h-l)}{m^*} = \frac{h-l}{m}$. That is, when the transparencies are superimposed with perfect alignment, our scheme recovers the original image with the same contrast as the conventional VSS scheme.

When $1 \le x \le m-1$ and $1 \le y \le N_y$, we obtain $\overline{a}^{(x,y)}$ as following:

$$\overline{a}^{(x,y)} \cdot m^* = \overline{H(V^1)} - \overline{H(V^0)} = \overline{H}(\begin{bmatrix} V^1[1] \\ \mathbf{L} \\ V^1[N_y] \end{bmatrix}) - \overline{H}(\begin{bmatrix} V^0[1] \\ \mathbf{L} \\ V^0[N_y] \end{bmatrix}) = \begin{bmatrix} \overline{a}_1 \cdot m_1 \\ \underline{\quad}\mathbf{L} \\ a_{N_y} \cdot m_{N_y} \end{bmatrix} = \overline{h} - \overline{l}$$

The contrast $\overline{a}^{(x,y)}$ refers to the average difference in weight in the combined share between a white pixel and a black pixel in the secret image, here $\overline{a}^{(x,y)} = (\overline{h} - \overline{l}) / m^*$. Namely, when there is a misalignment of $x$ pixels and $y$ pixels between the shares in horizontal and vertical direction, our scheme can still reveal the secret with a lower average contrast in the result image.

Since each block (with size $m^*$) for each share from matrix $B_0^*$ (resp. $B_1^*$) contains $N_y$ identical row vectors, and each row vector contains $N_x$ same row vectors of $B_0$ ( resp. $B_1$) which has the $m$ elements, there exist $(N_x \cdot m)!$ random column permutation methods to $C_0^*$ and $C_1^*$, respectively. To obtain the optimal contrast of the reconstructed secret image, we give a restricted column permutation method: the number of random column permutation of $C_0^*$ and $C_1^*$ is $m!$, not $(N_x \cdot m)!$. That is, order / sequence of column permutation of each of the $N_x$ same row vectors are the same as that of a row vector of $B_0$ and $B_1$. (For all kinds of three permutation methods, see more detail in Appendix C)

We give Example 5 to illustrate the definition and the permutation principle.

**Example 5**. A shift tolerant (2, 3)-VSS scheme (continuation of the example 1)

Based on an existing (2, 3)-VSS scheme, we construct a (2, 3)-VSS scheme that can tolerate pixels deviation in the horizontal direction when aligning the transparencies.

The basis matrices $B_0$ and $B_1$ of a (2, 3)-VSS scheme are $B_0 = \begin{bmatrix} 0 & 1 & 1 \\ 0 & 1 & 1 \\ 0 & 1 & 1 \end{bmatrix}$ and $B_1 = \begin{bmatrix} 0 & 1 & 1 \\ 1 & 1 & 0 \\ 1 & 0 & 1 \end{bmatrix}$.

The $i$th row of $B_0$ (resp. $B_1$), denoted by $m$-vector $B_0[i]$ (resp. $B_1[i]$), is distributed to the $i$-th participant as the $i$-th share, namely

$$B_0[1] = [0 \;\; 1 \;\; 1], \quad B_0[2] = [0 \;\; 1 \;\; 1], \quad B_0[3] = [0 \;\; 1 \;\; 1],$$

$$B_1[1] = [0 \;\; 1 \;\; 1], \quad B_1[2] = [1 \;\; 1 \;\; 0], \quad B_1[3] = [1 \;\; 0 \;\; 1].$$

We concatenate each row vector $B_0[i]$ (resp. $B_1[i]$) with itself and form a row vector twice as long. The symbol "$\mathbf{o}$" represents concatenation operation. For example,

$$B_0[1] \,\mathbf{o}\, B_0[1] = [0 \;\; 1 \;\; 1 \;\; 0 \;\; 1 \;\; 1], \quad B_1[1] \,\mathbf{o}\, B_1[1] = [0 \;\; 1 \;\; 1 \;\; 0 \;\; 1 \;\; 1].$$

The new basis matrices $B_0^*$ and $B_1^*$ are

$$B_0^* = \left[\begin{array}{ccc|ccc} 0 & 1 & 1 & 0 & 1 & 1 \\ 0 & 1 & 1 & 0 & 1 & 1 \\ 0 & 1 & 1 & 0 & 1 & 1 \end{array}\right] \begin{array}{l} \to 1 \\ \to 2 \\ \to 3 \end{array} \qquad B_1^* = \left[\begin{array}{ccc|ccc} 0 & 1 & 1 & 0 & 1 & 1 \\ 1 & 1 & 0 & 1 & 1 & 0 \\ 1 & 0 & 1 & 1 & 0 & 1 \end{array}\right] \begin{array}{l} \to 1 \\ \to 2 \\ \to 3 \end{array}$$

It is easily verified that $H(B_0^*[i]) = H(B_1^*[i]) = 4$ .That is, that the $i$th row vectors of $B_0^*$ and $B_1^*$ are indistinguishable, here $i \in \{1,2,3\}$. So the security condition is satisfied.

Let $C_0^*$ and $C_1^*$ be the collection of all matrices obtained by randomly permuting three columns on the left side of marked symbol " | " and the three columns on the right side with the same sequence of permutation, this is our permutation principle. $C_0^*$ and $C_1^*$ include 6 matrices, respectively. We list full columns permutation of $C_0^*$ and $C_1^*$ as follows (duplicated matrices will not be displayed):

$$C_0^* = \{ \left[\begin{array}{ccc|ccc} 0 & 1 & 1 & 0 & 1 & 1 \\ 0 & 1 & 1 & 0 & 1 & 1 \\ 0 & 1 & 1 & 0 & 1 & 1 \end{array}\right], \left[\begin{array}{ccc|ccc} 1 & 0 & 1 & 1 & 0 & 1 \\ 1 & 0 & 1 & 1 & 0 & 1 \\ 1 & 0 & 1 & 1 & 0 & 1 \end{array}\right], \left[\begin{array}{ccc|ccc} 1 & 1 & 0 & 1 & 1 & 0 \\ 1 & 1 & 0 & 1 & 1 & 0 \\ 1 & 1 & 0 & 1 & 1 & 1 \end{array}\right] \}$$

$$C_1^* = \{ \left[\begin{array}{ccc|ccc} 0 & 1 & 1 & 0 & 1 & 1 \\ 1 & 1 & 0 & 1 & 1 & 0 \\ 1 & 0 & 1 & 1 & 0 & 1 \end{array}\right], \left[\begin{array}{ccc|ccc} 0 & 1 & 1 & 0 & 1 & 1 \\ 1 & 0 & 1 & 1 & 0 & 1 \\ 1 & 1 & 0 & 1 & 1 & 0 \end{array}\right], \left[\begin{array}{ccc|ccc} 1 & 1 & 0 & 1 & 1 & 0 \\ 0 & 1 & 1 & 0 & 1 & 1 \\ 1 & 0 & 1 & 1 & 0 & 1 \end{array}\right],$$

$$\left[\begin{array}{ccc|ccc} 1 & 1 & 0 & 1 & 1 & 0 \\ 1 & 0 & 1 & 1 & 0 & 1 \\ 0 & 1 & 1 & 0 & 1 & 1 \end{array}\right], \left[\begin{array}{ccc|ccc} 1 & 0 & 1 & 1 & 0 & 1 \\ 0 & 1 & 1 & 0 & 1 & 1 \\ 1 & 1 & 0 & 1 & 1 & 0 \end{array}\right], \left[\begin{array}{ccc|ccc} 1 & 0 & 1 & 1 & 0 & 1 \\ 1 & 1 & 0 & 1 & 1 & 0 \\ 0 & 1 & 1 & 0 & 1 & 1 \end{array}\right] \}.$$

More detailed analysis of optimal contrast for a (2, 3)-STVSS scheme in the horizontal direction is given in Appendix C. An example to compute contrast of (2, 3)-STVSS scheme is given Appendix D.

### *3.2 The proposed shift tolerant (2, n)-VSS scheme*

Let $B_0$ and $B_1$ be the basis matrices for a $(2,n)$-VSS scheme with pixel expansion *m* and contrast $\alpha$. Similar to Liu's analysis [4], any two row vectors of the basis matrices of (2, *n*)-VSS scheme, such as *i th* row and *j th* row, can also be represented in the following form:

$$\begin{bmatrix} B_0[i] \\ B_0[j] \end{bmatrix} = \begin{bmatrix} 1\cdots1 & 0\cdots0 & 1\cdots1 & 0\cdots0 & 1\cdots1 & 0\cdots0 \\ \underbrace{1\cdots1}_{a'} & \underbrace{0\cdots0}_{b'} & \underbrace{0\cdots0}_{c} & \underbrace{1\cdots1}_{d} & \underbrace{1\cdots1}_{e} & \underbrace{0\cdots0}_{e} \end{bmatrix} \quad (1)$$

$$\begin{bmatrix} B_1[i] \\ B_1[j] \end{bmatrix} = \begin{bmatrix} 1\cdots1 & 0\cdots0 & 1\cdots1 & 0\cdots0 & 1\cdots1 & 0\cdots0 \\ \underbrace{1\cdots1}_{a'} & \underbrace{0\cdots0}_{b'} & \underbrace{0\cdots0}_{c} & \underbrace{1\cdots1}_{d} & \underbrace{0\cdots0}_{e} & \underbrace{1\cdots1}_{e} \end{bmatrix} \quad (2)$$

In the form above, some parameters of $(2,n)$-VSS scheme can be written as below:

$$m = a' + b' + c + d + 2e,\ \ l = a' + c + d + e,\ h = a' + c + d + 2e,\ \ \alpha = \frac{h-l}{m} = \frac{e}{m},\ i \neq j, i, j \in \{1, \ldots, n\}.$$

For simplicity, we use *m*-vector $B_0[i]$ ( resp. $B_1[i]$ ) to represent the *i*-th row of $B_0$ ( resp. $B_1$ ) is distributed to the *i*-th participant as the *i*-th share. $B_0[i] \circ B_0[i]$ be the concatenation of the two row vectors.

Next we will give the construction of (2, *n*)-STVSS, when any two shares are stacked precisely, or one share is shifted left (or right) by *x* ( $0 \le x \le m$ ) pixels position and shifted up (or down) by *y* （ $0 \le y < N_y$ ） pixels position relative to other share, we can still visually recover the secret image.

**Construction 1:** Construct (2, *n*)-STVSS based on (2, *n*)-VSS scheme.

**Input:** basis matrices $B_0$ and $B_1$ of (2, *n*)-VSS

**Output:** basis matrices $B_0^*$ and $B_1^*$ of (2, *n*)-STVSS

**Construct procedure**:

1. Construct block:

   for *i*=1 to *n*

$$Block_0^i = \left.\begin{bmatrix} B_0[i] \circ \cdots \circ B_0[i] \\ \vdots \quad \vdots \quad \vdots \\ \underbrace{B_0[i] \circ \cdots \circ B_0[i]}_{N_x} \end{bmatrix}\right\} N_y, \quad Block_1^i = \left.\begin{bmatrix} B_1[i] \circ \cdots \circ B_1[i] \\ \vdots \quad \vdots \quad \vdots \\ \underbrace{B_1[i] \circ \cdots \circ B_1[i]}_{N_x} \end{bmatrix}\right\} N_y$$

   end

2. Construct basis matrices $B_0^*$ and $B_1^*$:

$$B_0^* = \begin{bmatrix} Block_0^1 \\ \cdots \\ Block_0^i \\ \cdots \\ Block_0^n \end{bmatrix} \begin{matrix} \to 1 \\ \cdots \\ \to i \\ \cdots \\ \to n \end{matrix}, \quad B_1^* = \begin{bmatrix} Block_1^1 \\ \cdots \\ Block_1^i \\ \cdots \\ Block_1^n \end{bmatrix} \begin{matrix} \to 1 \\ \cdots \\ \to i \\ \cdots \\ \to n \end{matrix}$$

In construction 1, pixel expansion in vertical and horizontal are $m_x = N_x m$ and $m_y = N_y$, thus the total pixel expansion is $m^* = m_x m_y = N_x N_y m$.

Now we show the scheme satisfy security condition.

For any $i \in \{1, \cdots, n\}$, Hamming weight of $B_0^*[i]$ and $B_1^*[i]$ are:

$$H(B_0^*[i]) = \begin{pmatrix} H(B_0[i] \circ \cdots \circ B_0[i]) \\ \vdots \\ H(B_0[i] \circ \cdots \circ B_0[i]) \end{pmatrix}, H(B_1^*[i]) = \begin{pmatrix} H(B_1[i] \circ \cdots \circ B_1[i]) \\ \vdots \\ H(B_1[i] \circ \cdots \circ B_1[i]) \end{pmatrix}.$$

Since basis matrices $B_0$ and $B_1$ satisfy security condition, that is $H(B_0[i]) = H(B_1[i])$. Thus $H(B_0^*[i]) = H(B_1^*[i])$, this means the ***i***-th row vector of $B_0^*$ and $B_1^*$ are indistinguishable. So the security condition is satisfied.

**Theorem 2:** Construction 1 above is a 2 out of *n* shift tolerant visual cryptography scheme. The pixel expansion is $m^* = N_x N_y m$, when one share shifted pixels *x* pixels position in horizontal direction and shifted *y* pixels position in vertical direction according to the other share, the average contrast is

$$\overline{a}^{(x,y)} = -\frac{(N_x m - x)(N_y - y)}{N_x N_y m(m-1)} \cdot a \ , 1 \le x \le m-1 \, , 0 \le y < N_y \, .$$

**Proof:** In $(2, n)$-VSS scheme, any two out of $n$ shares are stacked to reconstruct the secret image. We suppose the two shares are share *i* (denoted by $S_i$) and share *j* (denoted by $S_j$). We now compute the average contrast when $S_j$ is shifted to left by *x* pixels and is shifted down by *y* pixels relative to $S_i$, here $1 \le x \le m-1$ and $0 \le y < N_y$. Without loss of generality, we analyze the stacking result of *i*th and *j*th row from basis matrices $B_0^*$ and $B_1^*$, denoted as $B_0^{*[i,j]} = \begin{bmatrix} Block_0^i \\ Block_0^j \end{bmatrix}$ and $B_1^{*[i,j]} = \begin{bmatrix} Block_1^i \\ Block_1^j \end{bmatrix}$, here $i \ne j, i, j \in \{1, \cdots, n\}$.

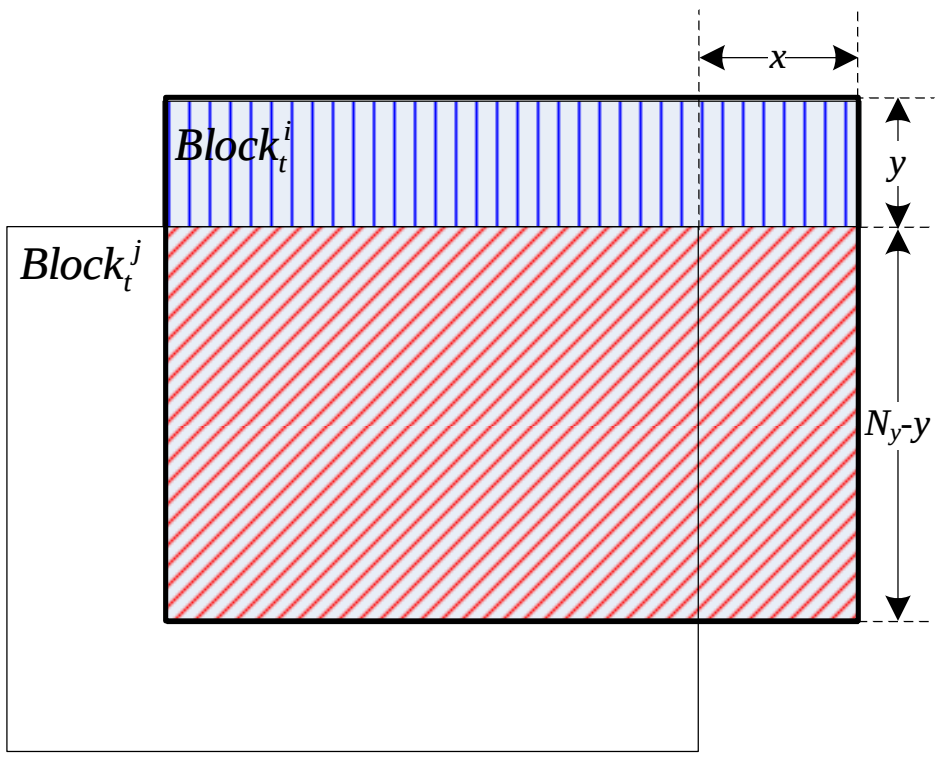

**Fig.**2 Reconstruction when $Block_t^j$ is shifted to left by *x* pixels and is shifted down by *y* pixels relative to $Block_t^i$, where $t = 0, 1$.

To get the average contrast, we need to compute the average Hamming weight $\bar{l}$ when $Block_0^i$ and $Block_0^j$ are stacked and average Hamming weight $\bar{h}$ when $Block_1^i$ and $Block_1^j$ are stacked (see Fig.2). This can be computed by two parts: (1) $\bar{l}_y$ and $\bar{h}_y$, stacking result of the top $y$ row vectors, i.e. the "" part in Fig.2; (2) $\bar{l}_{N_y-y}$ and $\bar{h}_{N_y-y}$, stacking result of the bottom $N_y - y$ row vectors, i.e. the "" part in Fig.3. Then $\bar{l} = \bar{l}_y + \bar{l}_{N_y-y}$ , $\bar{h} = \bar{h}_y + \bar{h}_{N_y-y}$ and the average contrast $\bar{a} = \dfrac{\bar{h} - \bar{l}}{m^*}$.

(i) *Part 1: compute the stacking result of the top y row vectors, i.e.* $\bar{l}_y$ *and* $\bar{h}_y$

Since $S_j$ is shifted down by $y$ pixels in vertical, that means $y$ rows of $Block_0^j$ or $Block_1^j$ are shifted out and another $y$ rows are shifted in, and let $\left.\begin{bmatrix} \tilde{B}_t[j] \circ \cdots \circ \tilde{B}_t[j] \\ \cdots \\ \tilde{B}_t[j] \circ \cdots \circ \tilde{B}_t[j] \end{bmatrix}\right\} y$ be $y$ row vectors that are shifted in, here $\tilde{B}_t[j]$ is different column permutation of $B_t[j], t \in \{0, 1\}, t = 0, 1$.

We now compute the average Hamming weight of top $y$ row vectors of stacking $Block_0^i$ and shifted $Block_0^j$, and stacking $Block_1^i$ and shifted $Block_1^j$, i.e.

$$\bar{l}_y = \overline{H}\left( \left.\begin{bmatrix} B_0[i] \circ \cdots \circ B_0[i] \\ \cdots \\ B_0[i] \circ \cdots \circ B_0[i] \end{bmatrix}\right\} y + \left.\begin{bmatrix} \tilde{B}_t[j] \circ \cdots \circ \tilde{B}_t[j] \\ \cdots \\ \tilde{B}_t[j] \circ \cdots \circ \tilde{B}_t[j] \end{bmatrix}\right\} y \right),$$

$$\bar{h}_y = \overline{H}\left( \left.\begin{bmatrix} B_1[i] \circ \cdots \circ B_1[i] \\ \cdots \\ B_1[i] \circ \cdots \circ B_1[i] \end{bmatrix}\right\} y + \left.\begin{bmatrix} \tilde{B}_t[j] \circ \cdots \circ \tilde{B}_t[j] \\ \cdots \\ \tilde{B}_t[j] \circ \cdots \circ \tilde{B}_t[j] \end{bmatrix}\right\} y \right).$$

We firstly compute $\overline{H}(B_0[i] + \tilde{B}_t[j]))$ and $\overline{H}(B_1[i] + \tilde{B}_t[j]))$ as below.

From formula（1）and (2), we give a simple representation of formula (1) and (2)

$$\begin{bmatrix} B_0[i] \\ B_0[j] \end{bmatrix} = \begin{bmatrix} \overbrace{1\cdots1}\;\overbrace{1\cdots1}\;\overbrace{1\cdots1}\;\overbrace{0\cdots0}\;\overbrace{0\cdots0}\;\overbrace{0\cdots0} \\ \underbrace{1\cdots1}_{a'}\;\underbrace{1\cdots1}_{d}\;\underbrace{1\cdots1}_{e}\;\underbrace{0\cdots0}_{b'}\;\underbrace{0\cdots0}_{c}\;\underbrace{0\cdots0}_{e} \end{bmatrix}, \quad \begin{bmatrix} B_1[i] \\ B_1[j] \end{bmatrix} = \begin{bmatrix} \overbrace{1\cdots1}\;\overbrace{1\cdots1}\;\overbrace{1\cdots1}\;\overbrace{0\cdots0}\;\overbrace{0\cdots0}\;\overbrace{0\cdots0} \\ \underbrace{1\cdots1}_{a'}\;\underbrace{1\cdots1}_{d}\;\underbrace{1\cdots1}_{e}\;\underbrace{0\cdots0}_{b'}\;\underbrace{0\cdots0}_{c}\;\underbrace{0\cdots0}_{e} \end{bmatrix}$$

It is easy verify that the number of 1's in $B_0[i]$ and $B_1[i]$ are both $a' + c + e$, and the number of 1 's in $B_0[j]$ and $B_1[j]$ are both $a' + d + e$.

Then we list the range of the number of 1's in $B_0[i] + \tilde{B}_t[j]$ and $B_1[i] + \tilde{B}_t[j]$ as below:

| $B_0[i]+\tilde{B}_t[j]$ | The minimum number of 1's( $h_0$ ) | The maximum number of 1's( $H_0$ ) |
|---|---|---|
| $B_0[i]+\tilde{B}_0[j]$ | $a'+\max(c,d)+e$ | $a'+\min(a',b')+c+d+2e$ |
| $B_0[i]+\tilde{B}_1[j]$ | $a'+\max(c,d)+e$ | $a'+\min(a',b')+c+d+2e$ |

and

| $B_1[i]+\tilde{B}_t[j]$ | The minimum number of 1's( $h_0$ ) | The maximum number of 1's( $H_0$ ) |
|---|---|---|
| $B_1[i]+\tilde{B}_0[j]$ | $a'+\max(c,d)+e$ | $a'+\min(a',b')+c+d+2e$ |
| $B_1[i]+\tilde{B}_1[j]$ | $a'+\max(c,d)+e$ | $a'+\min(a',b')+c+d+2e$ |

Suppose the OR result of $B_0[i]$ and $\tilde{B}_t[j]$ has the average Hamming weight $\bar{l}_1$, that is

$$\bar{l}_1=\overline{H}(\ B_0[i]+\tilde{B}_t[j])\ )=\frac{1}{2}\sum_{i'=h_0}^{H_0} i'\cdot p(H(B_0[i]\circ\tilde{B}_0[j])=i')+\frac{1}{2}\sum_{j'=h_0}^{H_0} j'\cdot p(H(B_0[i]\circ\tilde{B}_1[j])=j') \qquad (3)$$

Suppose the OR result of $B_1[i]$ and $\tilde{B}_t[j]$ has the average Hamming weight $\bar{h}_1$, that is

$$\bar{h}_1=\overline{H}(\ B_1[i]+\tilde{B}_t[j])\ )=\frac{1}{2}\sum_{i'=h_0}^{H_0} i'\cdot p(H(B_1[i]\circ\tilde{B}_0[j])=i')+\frac{1}{2}\sum_{j'=h_0}^{H_0} j'\cdot p(H(B_1[i]\circ\tilde{B}_1[j])=j') \qquad (4)$$

Since the matrices $B_0[i]$ are independent of the basis matrices $B_1[i]$, and $\tilde{B}_0[j]$ (resp. $\tilde{B}_1[j]$) are randomly column permutation of matrices $B_0[j]$ (resp. $B_1[j]$), from formula (3) and (4), we get $\bar{l}_1=\bar{h}_1$. Otherwise, it will contradict the property of basis matrices.

According to definition 2 and defined permutation principle, we get $\bar{l}_y=\bar{h}_y$.

(ii) *Part 2: compute the stacking result of the bottom* $N_y-y$ *row vectors, i.e.* $\bar{l}_{N_y-y}$ *and* $\bar{h}_{N_y-y}$.

By construction 1 above, every row of $Block_t^i$ in matrix $B_t^*$ is the same, here $t$=0, 1. Thus to compute the stacking result of $Block_t^i$ and $Block_t^j$, we can just analyze the stacking of $[B_t[i]\circ\cdots\circ B_t[i]]$ and $[B_t[j]\circ\cdots\circ B_t[j]]$.

We create the matrix collection $M_t[X]$ by using the two row vectors $[B_t[i]\circ\cdots\circ B_t[i]]$ and $[B_t[j]\circ\cdots\circ B_t[j]]$. That is

$$M_t[X]=\begin{bmatrix} \overbrace{\qquad\qquad\qquad}^{N_x} \\ B_t[i]\circ\cdots\circ B_t[i] \\ B_t[j]\circ\cdots\circ B_t[j] \end{bmatrix}, \text{ where } X=\{i,j\}.$$

By (1), (2), $M_t[X]$ has following two representation form respectively.

$$M_0[X]=\begin{bmatrix} \overbrace{\qquad\qquad\qquad\qquad\qquad\qquad\qquad\qquad\qquad\qquad\qquad\qquad\qquad\qquad\qquad\qquad}^{N_x} \\ 1\cdots1\ 0\cdots0\ 1\cdots1\ 0\cdots0\ 1\cdots1\ 0\cdots0\ \cdots\ 1\cdots1\ 0\cdots0\ 1\cdots1\ 0\cdots0\ 1\cdots1\ 0\cdots0 \\ \underbrace{1\cdots1}_{a'}\ \underbrace{0\cdots0}_{b'}\ \underbrace{0\cdots0}_{c}\ \underbrace{1\cdots1}_{d}\ \underbrace{1\cdots1}_{e}\ \underbrace{0\cdots0}_{e}\ \cdots\ \underbrace{1\cdots1}_{a'}\ \underbrace{0\cdots0}_{b'}\ \underbrace{0\cdots0}_{c}\ \underbrace{1\cdots1}_{d}\ \underbrace{1\cdots1}_{e}\ \underbrace{0\cdots0}_{e} \end{bmatrix}$$

$$M_1[X]=\begin{bmatrix} \overbrace{1\cdots10\cdots0\,1\cdots1\,0\cdots01\cdots10\cdots0\cdots1\cdots10\cdots01\cdots10\cdots01\cdots10\cdots0}^{N_x} \\ \underbrace{1\cdots1}_{a'}\underbrace{0\cdots0}_{b'}\underbrace{0\cdots0}_{c}\underbrace{1\cdots1}_{d}\underbrace{0\cdots0}_{e}\underbrace{1\cdots1}_{e}\cdots\underbrace{1\cdots1}_{a'}\underbrace{0\cdots0}_{b'}\underbrace{0\cdots0}_{c}\underbrace{1\cdots1}_{d}\underbrace{0\cdots0}_{e}\underbrace{1\cdots1}_{e} \end{bmatrix}$$

Let $C_t[X]$ be the collection of matrices by permuting $M_t[X]$ according to the permutation principle of the definition 2 above in our permutation way, thus $C_t[X]$ has $m!$ matrices, not be $(N_x m)!$.

Shift the second row ($j$ row) of the $m!$ share matrices in $C_t[X]$ to left (or right) by $x$ pixels, and let $c_1\cdots c_x$ be the $x$-bit string that is shifted in , where $c_i\in\{0,1\}, i=1,\cdots,x$ .When $[B_t[j]\circ\cdots\circ B_t[j]]$ is shifted to left by $x$ pixels relative to $[B_t[i]\circ\cdots\circ B_t[i]]$, then the shifted matrix is denoted by $M_t^{(x)}[X]$,

$$M_0^{(x)}[X]=\begin{bmatrix} \overbrace{*\cdots*1\cdots10\cdots01\cdots10\cdots01\cdots10\cdots0}^{N_x-1}\quad\cdots\quad 1\cdots10\cdots01\cdots10\cdots01\cdots10\cdots0 \\ \underbrace{1\cdots1}_{a'}\underbrace{0\cdots0}_{b'}\underbrace{0\cdots0}_{c}\underbrace{1\cdots1}_{d}\underbrace{1\cdots1}_{e}\underbrace{0\cdots0}_{e}\quad\cdots\quad\underbrace{1\cdots1}_{a'}\underbrace{0\cdots0}_{b'}\underbrace{0\cdots0}_{c}\underbrace{1\cdots1}_{d}\underbrace{1\cdots1}_{e}\underbrace{0\cdots0}_{e}\underbrace{c_1\cdots c_x}_{x} \end{bmatrix}$$

$$M_1^{(x)}[X]=\begin{bmatrix} \overbrace{*\cdots*1\cdots10\cdots01\cdots10\cdots01\cdots10\cdots0}^{N_x-1}\quad\cdots\quad 1\cdots10\cdots01\cdots10\cdots01\cdots10\cdots0 \\ \underbrace{1\cdots1}_{a'}\underbrace{0\cdots0}_{b'}\underbrace{0\cdots0}_{c}\underbrace{1\cdots1}_{d}\underbrace{0\cdots0}_{e}\underbrace{1\cdots1}_{e}\quad\cdots\quad\underbrace{1\cdots1}_{a'}\underbrace{0\cdots0}_{b'}\underbrace{0\cdots0}_{c}\underbrace{1\cdots1}_{d}\underbrace{0\cdots0}_{e}\underbrace{1\cdots1}_{e}\underbrace{c_1\cdots c_x}_{x} \end{bmatrix}.$$

And the collection of matrices by permuting $M_t^{(x)}[X]$ in our permutation way will be denoted by $C_t^{(x)}[X]$.

Now we begin to compute the contrast of the recovered secret image when there is $x$ pixels deviation between these two rows.

According to results of Ref. [4], the total Hamming weight of the stacking of the shifted two rows can be calculated by the total number of the 1's subtract the number of the 1's that are ineffective. There are three cases when a black subpixel 1 is in effective: (1) when that is in the top right corner of the matrix $M_t^{(x)}[X]$ and the corresponding position has 1 shifted in; (2) when that is in the bottom left corner of the matrix which is shifted out; (3) when an overlap happens after a shift.

Let $p_{c_1\cdots c_x}$ denote the probability that a string $c_1\cdots c_x$ is shifted in, and we have $\sum_{c_1\cdots c_x\in\{0,1\}} p_{c_1\cdots c_x}=1$.

In the first case, let $s^1_{0,c_1\cdots c_x}$ and $s^1_{1,c_1\cdots c_x}$ denote the number of 1's that are ineffective for the collections $C_0^{(x)}[X]$ and $C_1^{(x)}[X]$ respectively when $c_1\cdots c_x$ is shifted in. Suppose the Hamming weight of $c_1\cdots c_x$ is $s$, i.e. $\mathrm{H}(c_1\cdots c_x)=s$. The total number of 1's that are ineffective in the top right corner of all the matrices in $C_0^{(x)}[X]$ and $C_1^{(x)}[X]$ is $s^1_{0,c_1\cdots c_x} = s^1_{1,c_1\cdots c_x} = s\cdot\frac{a'+c+e}{m}m!$;

In the second case, let $s^2_{0,c_1\cdots c_x}$ and $s^2_{1,c_1\cdots c_x}$ denote the number of 1's that are ineffective for the collections $C_0^{(x)}[X]$ and $C_1^{(x)}[X]$ respectively when $c_1\cdots c_x$ is shifted in. The total number of 1's that are ineffective in the top right corner of all the matrices in $C_0^{(x)}[X]$ and $C_1^{(x)}[X]$ is $s^2_{0,c_1\cdots c_x} = s^2_{1,c_1\cdots c_x} = x\cdot\frac{a'+d+e}{m}m!$;

In the third case, let $s^3_{0,c_1\cdots c_x}$ and $s^3_{1,c_1\cdots c_x}$ denote the number of 1's that are ineffective for the collections $C_0^{(x)}[X]$ and $C_1^{(x)}[X]$ respectively when $c_1\cdots c_x$ is shifted in. Pattern $\begin{bmatrix}1\\1\end{bmatrix}$ in the shifted matrices is the shifted result of the following four patterns $\begin{bmatrix}1 & \cdots & 0\\ 0 & \underbrace{\cdots}_{x} & 1\end{bmatrix}$, $\begin{bmatrix}1 & \cdots & 0\\ 1 & \underbrace{\cdots}_{x} & 1\end{bmatrix}$, $\begin{bmatrix}1 & \cdots & 1\\ 0 & \underbrace{\cdots}_{x} & 1\end{bmatrix}$ and $\begin{bmatrix}1 & \cdots & 1\\ 1 & \underbrace{\cdots}_{x} & 1\end{bmatrix}$ in the collections $C_0^{(x)}[X]$ and $C_1^{(x)}[X]$. We calculate the probability of the pattern $\begin{bmatrix}1 & \cdots & 0\\ 0 & \underbrace{\cdots}_{x} & 1\end{bmatrix}$ in $C_1^{(x)}[X]$ for example, and the other three patterns can be calculated similarly. Since the basis matrices are composed by $N_x$ same blocks, we first compute the probability in the first block. The probability that the pattern $\begin{bmatrix}1\\0\end{bmatrix}$ appears at the column $j_1$ $(1\le j_1\le m)$ in the first block of the matrices of the collection $C_1^{(x)}[X]$ is $\frac{c+e}{m}$. Based on this, the probability that the pattern $\begin{bmatrix}0\\1\end{bmatrix}$ appears at the column ($j_1$+$x$) in the matrices of the collection $C_1^{(x)}[X]$ is $\frac{d+e}{m-1}$. So the probability that the pattern $\begin{bmatrix}1 & \cdots & 0\\ 0 & \underbrace{\cdots}_{x} & 1\end{bmatrix}$ appears both at column $j_1$ $(1\le j_1\le m)$ and $j_1$+$x$ in first and second block of the matrices of the collection $C_1^{(x)}[X]$ is $\frac{c+e}{m}\frac{d+e}{m-1}$.

The remaining three patterns can be compute in the same way.

| Collections\Patterns | $\begin{bmatrix}1 & \cdots & 0\\ 0 & \underbrace{\cdots}_{x} & 1\end{bmatrix}$ | $\begin{bmatrix}1 & \cdots & 0\\ 1 & \underbrace{\cdots}_{x} & 1\end{bmatrix}$ | $\begin{bmatrix}1 & \cdots & 1\\ 0 & \underbrace{\cdots}_{x} & 1\end{bmatrix}$ | $\begin{bmatrix}1 & \cdots & 1\\ 1 & \underbrace{\cdots}_{x} & 1\end{bmatrix}$ |
|---|---|---|---|---|
| $C_0^{(x)}[X]$ | $\frac{c}{m}\frac{d}{m-1}$ | $\frac{a'+e}{m}\frac{d}{m-1}$ | $\frac{c}{m}\frac{a'+e}{m-1}$ | $\frac{a'+e}{m}\frac{a'+e-1}{m-1}$ |
| $C_1^{(x)}[X]$ | $\frac{c+e}{m}\frac{d+e}{m-1}$ | $\frac{a'}{m}\frac{d+e}{m-1}$ | $\frac{c+e}{m}\frac{a'}{m-1}$ | $\frac{a'}{m}\frac{a'-1}{m-1}$ |

We notice that each row vector of $M_0^{(x)}[X]$ (resp. $M_1^{(x)}[X]$) contains $N_x$ identical row vectors of matrix $B_0$ (resp. $B_1$), we partition the $N_x \cdot m$ columns into two sections: the first ($N_x$-1) blocks and the last block.

In the first ($N_x$-1) blocks, there are $m$ choices for the value of $j_1$. The number of 1's that are overlapped in the collections $C_0^{(x)}[X]$ and $C_1^{(x)}[X]$ in the first ($N_x$-1) blocks are respectively

$$(N_x - 1)\left(\frac{c}{m}\frac{d}{m-1} + \frac{a'+e}{m}\frac{d}{m-1} + \frac{c}{m}\frac{a'+e}{m-1} + \frac{a'+e}{m}\frac{a'+e-1}{m-1}\right)mm!$$

and $$(N_x - 1)\left(\frac{c+e}{m}\frac{d+e}{m-1} + \frac{a'}{m}\frac{d+e}{m-1} + \frac{c+e}{m}\frac{a'}{m-1} + \frac{a'}{m}\frac{a'-1}{m-1}\right)mm!.$$

In the last block, there are $(m-x)$ choices for the value of $j_1$. The number of 1's that are overlapped in the collections $C_0^{(x)}[X]$ and $C_1^{(x)}[X]$ in the last block are respectively

$$\left(\frac{c}{m}\frac{d}{m-1} + \frac{a'+e}{m}\frac{d}{m-1} + \frac{c}{m}\frac{a'+e}{m-1} + \frac{a'+e}{m}\frac{a'+e-1}{m-1}\right)(m-x)m!$$

and $$\left(\frac{c+e}{m}\frac{d+e}{m-1} + \frac{a'}{m}\frac{d+e}{m-1} + \frac{c+e}{m}\frac{a'}{m-1} + \frac{a'}{m}\frac{a'-1}{m-1}\right)(m-x)m!.$$

So the total number of 1's that are overlapped in the collections $C_0^{(x)}[X]$ and $C_1^{(x)}[X]$ are

$$s^3_{0,c_1 \cdots c_x} = (N_x - 1)\left(\frac{c}{m}\frac{d}{m-1} + \frac{a'+e}{m}\frac{d}{m-1} + \frac{c}{m}\frac{a'+e}{m-1} + \frac{a'+e}{m}\frac{a'+e-1}{m-1}\right)mm! +$$
$$+\left(\frac{c}{m}\frac{d}{m-1} + \frac{a'+e}{m}\frac{d}{m-1} + \frac{c}{m}\frac{a'+e}{m-1} + \frac{a'+e}{m}\frac{a'+e-1}{m-1}\right)(m-x)m!$$

$$s^3_{1,c_1 \cdots c_x} = (N_x - 1)\left(\frac{c+e}{m}\frac{d+e}{m-1} + \frac{a'}{m}\frac{d+e}{m-1} + \frac{c+e}{m}\frac{a'}{m-1} + \frac{a'}{m}\frac{a'-1}{m-1}\right)(m-1)m!$$
$$+\left(\frac{c+e}{m}\frac{d+e}{m-1} + \frac{a'}{m}\frac{d+e}{m-1} + \frac{c+e}{m}\frac{a'}{m-1} + \frac{a'}{m}\frac{a'-1}{m-1}\right)(m-x)m!$$

Let $T_{0,c_1 \cdots c_x}$ and $T_{1,c_1 \cdots c_x}$ denote the total number of 1 in $N_y - y$ rows of $Block_t^i$ and $Block_t^j$ when a string of pixels $c_1 \cdots c_x$ is shifted in, since we have supposed $H(c_1 \cdots c_x) = s$, we have

$$T_{0,c_1 \cdots c_x} = T_{1,c_1 \cdots c_x} = (N_y - y)N_x(2a'+c+d+2e+s)m!.$$

Let $w_{0,c_1 \cdots c_x}$ (resp. $w_{1,c_1 \cdots c_x}$) denotes the total stacking Hamming weight of $N_y - y$ rows of stacking result of $Block_0^i$ and $Block_0^j$ (resp. $Block_1^i$ and $Block_1^j$) when a string $c_1 \cdots c_x$ are shifted in, we have

$$w_{0,c_1 \cdots c_x} = T_{0,c_1 \cdots c_x} - (N_y - y)s^1_{0,c_1 \cdots c_x} - (N_y - y)s^2_{0,c_1 \cdots c_x} - (N_y - y)s^3_{0,c_1 \cdots c_x}$$
$$w_{1,c_1 \cdots c_x} = T_{1,c_1 \cdots c_x} - (N_y - y)s^1_{1,c_1 \cdots c_x} - (N_y - y)s^2_{1,c_1 \cdots c_x} - (N_y - y)s^3_{1,c_1 \cdots c_x}.$$

Let $\bar{l}_{N_y-y}$ (resp. $\bar{h}_{N_y-y}$) denotes the average Hamming weight of $N_y - y$ rows of stacking result of $Block_0^i$ and $Block_0^j$ (resp. $Block_1^i$ and $Block_1^j$) when a string $c_1 \mathbf{L}\, c_x$ are shifted in, then

$$\bar{l}_{N_y-y} = \frac{\sum_{c_1 \mathbf{L} c_x \in \{0,1\}} w_{0,c_1 \mathbf{L} c_x} \cdot p_{c_1 \mathbf{L} c_x}}{m!} \text{ and } \bar{h}_{N_y-y} = \frac{\sum_{c_1 \mathbf{L} c_x \in \{0,1\}} w_{1,c_1 \mathbf{L} c_x} \cdot p_{c_1 \mathbf{L} c_x}}{m!}.$$

We have

$$\begin{aligned}
&\bar{h}_{N_y-y} - \bar{l}_{N_y-y} \\
&= \frac{\sum_{c_1 \mathbf{L} c_x \in \{0,1\}} w_{1,c_1 \mathbf{L} c_x} \cdot p_{c_1 \mathbf{L} c_x}}{m!} - \frac{\sum_{c_1 \mathbf{L} c_x \in \{0,1\}} w_{0,c_1 \mathbf{L} c_x} \cdot p_{c_1 \mathbf{L} c_x}}{m!} \\
&= \frac{(w_{1,c_1 \mathbf{L} c_x} - w_{0,c_1 \mathbf{L} c_x})\sum_{c_1 \mathbf{L} c_x \in \{0,1\}} p_{c_1 \mathbf{L} c_x}}{m!} = \frac{w_{1,c_1 \mathbf{L} c_x} - w_{0,c_1 \mathbf{L} c_x}}{m!} \\
&= ((T_{1,c_1 \mathbf{L} c_x} - (N_y - y)s^1_{1,c_1 \mathbf{L} c_x} - (N_y - y)s^2_{1,c_1 \mathbf{L} c_x} - (N_y - y)s^3_{1,c_1 \mathbf{L} c_x}) \\
&\quad - (T_{0,c_1 \mathbf{L} c_x} - (N_y - y)s^1_{0,c_1 \mathbf{L} c_x} - (N_y - y)s^2_{0,c_1 \mathbf{L} c_x} - (N_y - y)s^3_{0,c_1 \mathbf{L} c_x}))/m! \\
&= -\frac{e(N_x m - x)(N_y - y)}{m(m-1)}.
\end{aligned}$$

By combining the result of part 1 and part 2, the average contrast is

$$\begin{aligned}
\bar{a}^{(x,y)} &= \frac{\bar{h}_y - \bar{l}_y + \bar{h}_{N_y-y} - \bar{l}_{N_y-y}}{m^*} = \frac{-\dfrac{e(N_x m - x)(N_y - y)}{m(m-1)}}{N_x N_y m} \\
&= -\frac{(N_y - y)(N_x m - x)}{N_x N_y m(m-1)} \cdot \frac{(h-l)}{m} = -\frac{(N_x m - x)(N_y - y)}{N_x N_y m(m-1)} \cdot a. \quad \blacksquare
\end{aligned}$$

Using theorem 2, we obtain the following corollary 1.

**Corollary 1:** In Construction 1 above, when there is just $y$ pixels deviation in vertical direction, i.e. $x$=0, the average contrast is $\bar{a}^{(0,y)} = \dfrac{N_y - y}{N_y} \cdot a$ .

According to our construction 1, we now discuss the average contrast when there is $m$ pixels deviation in horizontal direction and $y$ pixels deviation in vertical direction.

**Theorem 3**: In construction 1 above, when there is $m$ pixels deviation in horizontal direction and $y$ pixels deviation in vertical direction, i.e. $x$=$m$, the average contrast is $\bar{a}^{(m,y)} = \dfrac{(N_x - 1)(N_y - y)}{N_x N_y} \cdot a$ .

**Proof:** For simply, we use the matrix collections and symbols of Theorem 2 to prove the Corollary 2. Similar to Theorem 2, the average contrast can be computed by two parts.

(i) Part 1: The first part is the same with the first part of Theorem 2 and we get

$$\bar{h}_y = \bar{l}_y . \tag{5}$$

(ii) Part 2: There are some differences with the the second part of Theorem 2.

Similar to the definition in Theorem 2,

$$M_t[X] = \begin{bmatrix} \overbrace{B_t[i] \circ \cdots \circ B_t[i]}^{N_x} \\ B_t[j] \circ \cdots \circ B_t[j] \end{bmatrix}, \text{ where } X=\{i, j\}.$$

When the $j$-th row is shifted to left by $x$ ( $x = m$) pixels according to $i$-th row, this means there is a block shifted in, let $\tilde{B}_t[j]$ denote the block shifted in, that is

$$M_t^{(m)}[X] = \left[\begin{array}{c|cccc} * & B_t[i] & \cdots & B_t[i] & B_t[i] \\ B_t[j] & B_t[j] & \cdots & B_t[j] & \tilde{B}_t[j] \end{array}\right]$$

According to security condition of $B_0$ and $B_1$, $H(B_0[i]) = H(B_1[i]) = H(B_0[j]) = H(B_1[j])$.

Suppose the OR result of $B_0[i]$ and $\tilde{B}_t[j]$ has the average Hamming weight $\bar{l}_1$, that is

$$\bar{l}_1 = H(B_0[i] + \tilde{B}_t[j]) .$$

Suppose the OR result of $B_1[i]$ and $\tilde{B}_t[j]$ has the average Hamming weight $\bar{h}_1$, that is

$$\bar{h}_1 = H(B_1[i] + \tilde{B}_t[j]) .$$

Similar to the proof of Step 1 of Theorem 2, we can obtain

$$\bar{l}_1 = \bar{h}_1 . \tag{6}$$

Since $H(B_0[i] + B_0[j]) = l$, the average Hamming weight by stacking two rows of all the matrices in $C_0^{(m)}[X]$ is $\overline{H}(C_0^{(m)}[i] + C_0^{(m)}[j]) = (N_x - 1) \cdot l + \bar{l}_1$.

Since $H(B_1[i] + B_1[j]) = h$, the average Hamming weight by stacking two rows of all the matrices in $C_1^{(m)}[X]$ is $\overline{H}(C_1^{(m)}[i] + C_1^{(m)}[j]) = (N_x - 1) \cdot h + \bar{h}_1$.

Let $\bar{l}_{N_y - y}$ (resp. $\bar{h}_{N_y - y}$) denotes the average Hamming weight of $N_y - y$ rows of stacking result of $Block_0^i$ and $Block_0^j$ (resp. $Block_1^i$ and $Block_1^j$) when $\tilde{B}_t[j]$ is shifted in, then

$$\bar{l}_{N_y - y} = (N_y - y) \cdot \overline{H}(C_0^{(m)}[i] + C_0^{(m)}[j]) = (N_y - y) \cdot ((N_x - 1) \cdot l + \bar{l}_1) . \tag{7}$$

$$\bar{h}_{N_y - y} = (N_y - y) \cdot \overline{H}(C_1^{(m)}[i] + C_1^{(m)}[j]) = (N_y - y) \cdot ((N_x - 1) \cdot h + \bar{h}_1) . \tag{8}$$

By (5), (6), (7), and (8), thus the average contrast is

$$\bar{a}^{(m,y)} = \frac{\bar{h}_{N_y - y} - \bar{l}_{N_y - y} + \bar{h}_y - \bar{l}_y}{m^*} = \frac{(N_y - y)[((N_x - 1) \cdot h + \bar{h}_1) - ((N_x - 1) \cdot l + \bar{l}_1)]}{N_x N_y m}$$

$$= \frac{(N_y - y)(N_x - 1)(h - l)}{N_x N_y m} = \frac{(N_y - y)(N_x - 1)}{N_x N_y} \cdot a \; . \qquad \blacksquare$$

We shall give a shift tolerant (2, 3)-VSS scheme to illustrate the theorem above.

**Example 6.** A shift tolerant (2, 3)-VSS scheme in both horizontal and vertical direction.

$B_0$ and $B_1$ are basis matrices of a (2, 3)-VSS scheme ( see Example 1):

$$B_0=\begin{bmatrix}0&1&1\\0&1&1\\0&1&1\end{bmatrix}\qquad B_1=\begin{bmatrix}0&1&1\\1&1&0\\1&0&1\end{bmatrix}.$$

We have

$$B_0[1]=[0\ \ 1\ \ 1], B_0[2]=[0\ \ 1\ \ 1], B_0[3]=[0\ \ 1\ \ 1],$$

$$B_1[1]=[0\ \ 1\ \ 1], B_1[2]=[1\ \ 1\ \ 0], B_1[3]=[1\ \ 0\ \ 1].$$

According to Construction 1, we have for example:

$$Block_0^1=\begin{bmatrix}B_0[1]\circ B_0[1]\\B_0[1]\circ B_0[1]\end{bmatrix}=\begin{bmatrix}0&1&1&0&1&1\\0&1&1&0&1&1\end{bmatrix},\quad Block_1^1=\begin{bmatrix}B_1[1]\circ B_1[1]\\B_1[1]\circ B_1[1]\end{bmatrix}=\begin{bmatrix}0&1&1&0&1&1\\0&1&1&0&1&1\end{bmatrix}.$$

By using similar method above, we get new matrices $B_0^*$ and $B_1^*$ as follows:

$$B_0^*=\left[\begin{array}{cccccc}0&1&1&0&1&1\\0&1&1&0&1&1\\ \hdashline 0&1&1&0&1&1\\0&1&1&0&1&1\\ \hdashline 0&1&1&0&1&1\\0&1&1&0&1&1\end{array}\right]\begin{array}{l}\}\to 1\\ \\ \}\to 2\\ \\ \}\to 3\\ \end{array}\qquad B_1^*=\left[\begin{array}{cccccc}0&1&1&0&1&1\\0&1&1&0&1&1\\ \hdashline 1&1&0&1&1&0\\1&1&0&1&1&0\\ \hdashline 1&0&1&1&0&1\\1&0&1&1&0&1\end{array}\right]\begin{array}{l}\}\to 1\\ \\ \}\to 2\\ \\ \}\to 3\\ \end{array}$$

Suppose share 1 (denoted by $S_1$) and share 3 (denoted by $S_3$) are stacked to reconstruct the secret image. The original white (resp. black) pixel is reconstructed by stacking $B_0^*[1]$ and $B_0^*[3]$ (resp. $B_1^*[1]$ and $B_1^*[3]$). When $S_3$ is shifted to left by *x* pixels and up by *y* pixels facing onto $S_1$, there will be some pixels shifted in and some shifted out. Now we will give an example in Table 6 (resp. Table 7) to show the reconstruction process of the original white (resp. black) pixel.

**Table 6.** Reconstruction process of the original white pixel when there is diagonal deviation.

| | (i). $x$=0 | (ii). $x$=1, $y$=1 |
|---|---|---|
| $B_0^*[1]$ | $\begin{bmatrix}0&1&1&0&1&1\\0&1&1&0&1&1\end{bmatrix}$ | $\begin{bmatrix}0&1&1&0&1&1\\0&1&1&0&1&1\end{bmatrix}$ |
| $B_0^*[3]$ | $\begin{bmatrix}0&1&1&0&1&1\\0&1&1&0&1&1\end{bmatrix}$ | $\left[\begin{array}{c;{2pt/2pt}cccccc}0&1&1&0&1&1&1\\ \hdashline 0&1&1&0&1&1&1\\1&1&0&1&1&0&0\end{array}\right]$ |
| $V^0=B_0^*[1]+B_0^*[3]$ | $\begin{bmatrix}0&1&1&0&1&1\\0&1&1&0&1&1\end{bmatrix}$ | $\begin{bmatrix}1&1&1&1&1&1\\1&1&1&1&1&1\end{bmatrix}$ |
| $H(V^0)$ | 8 | 12 |

**Table 7.** Reconstruction process of the original black pixel when there is diagonal deviation.

| | (i). $x$=0 | (ii). $x$=1, $y$=1 |
|---|---|---|
| $B_1^*[1]$ | $\begin{bmatrix} 0 & 1 & 1 & 0 & 1 & 1 \\ 0 & 1 & 1 & 0 & 1 & 1 \end{bmatrix}$ | $\begin{bmatrix} 0 & 1 & 1 & 0 & 1 & 1 \\ 0 & 1 & 1 & 0 & 1 & 1 \end{bmatrix}$ |
| $B_1^*[3]$ | $\begin{bmatrix} 1 & 0 & 1 & 1 & 0 & 1 \\ 1 & 0 & 1 & 1 & 0 & 1 \end{bmatrix}$ | $\begin{bmatrix} 1 & 0 & 1 & 1 & 0 & 1 & 1 \\ 1 & 0 & 1 & 1 & 0 & 1 & 1 \\ 1 & 1 & 0 & 1 & 1 & 0 & 0 \end{bmatrix}$ |
| $V^1 = B_1^*[1] + B_1^*[3]$ | $\begin{bmatrix} 1 & 1 & 1 & 1 & 1 & 1 \\ 1 & 1 & 1 & 1 & 1 & 1 \end{bmatrix}$ | $\begin{bmatrix} 0 & 1 & 1 & 0 & 1 & 1 \\ 1 & 1 & 1 & 1 & 1 & 1 \end{bmatrix}$ |
| $H(V^1)$ | 12 | 10 |

Contrast can be compute in this way: for example, in situation (ii), $x$=1 and $y$=1, we have $H(V^0) = 12$ and $H(V^1) = 10$. Thus contrast of this situation is $\frac{H(V^1) - H(V^0)}{m^*} = \frac{10-12}{12} = -\frac{1}{6}$. Since there are $6 \cdot 6 \cdot 6 \cdot 6$ different cases, we have to compute the average contrast $\overline{a}^{(1,1)}$ by using algorithm (see algorithm in Appendix C). The result computed is $\overline{a}^{(1,1)} = -\frac{5}{72}$ .By theorem 2 above, the average contrast $\overline{a}^{(1,1)} = -\frac{(2.3-1)(2-1)}{2 \cdot 2 \cdot 3.2} \cdot \frac{1}{3} = -\frac{5}{72}$. The two results are same. The experiment results are given in Appendix F.

In theorem 2 above, when $N_x = N_y = 1$, the shift tolerant $(2, n)$-VSS scheme equals a conventional $(2, n)$-VSS, we obtain immediately the following the result from the Theorem 2.

**Corol1ary 2.** In conventional (2, $n$)-VSS scheme with pixel expansion $m$ and contrast $a$, when a share is shifted $x$ pixels position relative to other share in horizontal direction, the average contrast of the recovered image $\overline{a}^{(x,0)} = -\frac{(m-x)(h-l)}{m^2(m-1)} = -\frac{(m-x)}{m(m-1)} a$ , here $(1 \le x \le m-1)$

Obviously, when $n$ = 2, this scheme is equivalent to the (2, 2)-VSS scheme in [4], also see Theorem 1 above.

# 4. A general approach to construct shift tolerant (*k*, *n*)-VSS scheme

Using a procedure similar to the one discussed in previous sections, one can easily construct a more general shift tolerant ($k$, $n$)-VSS scheme for $3 \le k \le n$.

Let $B_0$ and $B_1$ be the basis matrices for a $(k, n)$-VCS with pixel expansion $m$ and contrast $a$.

**Construction 2:** Construct ($k$, $n$)-STVSS scheme based on a ($k$, $n$)-VSS scheme.

**Input:** the basis matrices $B_0$ and $B_1$ of a ($k$, $n$)-VSS scheme

**Output:** the basis matrices $B_0^*$ and $B_1^*$ of the ($k$, $n$)-STVSS scheme

**Construction procedure:**

1. Construct blocks:

for $i$=1 to $n$

$$Block_0^i = \underbrace{\begin{bmatrix} B_0[i] \circ \cdots \circ B_0[i] \\ \vdots \quad \vdots \quad \vdots \\ B_0[i] \circ \cdots \circ B_0[i] \end{bmatrix}}_{N_x} \Bigg\} N_y, \quad Block_1^i = \underbrace{\begin{bmatrix} B_1[i] \circ \cdots \circ B_1[i] \\ \vdots \quad \vdots \quad \vdots \\ B_1[i] \circ \cdots \circ B_1[i] \end{bmatrix}}_{N_x} \Bigg\} N_y$$

end

2. Construct basis matrices $B_0^*$ and $B_1^*$:

$$B_0^* = \begin{bmatrix} Block_0^1 \\ \cdots \\ Block_0^i \\ \cdots \\ Block_0^n \end{bmatrix} \begin{matrix} \to 1 \\ \cdots \\ \to i \\ \cdots \\ \to n \end{matrix}, \quad B_1^* = \begin{bmatrix} Block_1^1 \\ \cdots \\ Block_1^i \\ \cdots \\ Block_1^n \end{bmatrix} \begin{matrix} \to 1 \\ \cdots \\ \to i \\ \cdots \\ \to n \end{matrix}$$

This construction procedure is demonstrated below through an example (3, 4)-STVSS scheme.

**Example 7.** An example of (3, 4)-STVSS scheme.

Consider a (3, 4)-VSS scheme with basis matrices $B_0$ and $B_1$.

$$B_0 = \begin{bmatrix} 0 & 0 & 1 & 1 & 1 & 0 \\ 0 & 0 & 1 & 1 & 0 & 1 \\ 0 & 0 & 1 & 0 & 1 & 1 \\ 0 & 0 & 0 & 1 & 1 & 1 \end{bmatrix}, \quad B_0 = \begin{bmatrix} 1 & 0 & 0 & 0 & 1 & 1 \\ 0 & 1 & 0 & 0 & 1 & 1 \\ 0 & 0 & 1 & 0 & 1 & 1 \\ 0 & 0 & 0 & 1 & 1 & 1 \end{bmatrix}$$

We will construct a (3, 4)-STVSS scheme based on this (3, 4)-VSS scheme. Let $N_x = 2$ and $N_y = 2$, the basis matrices $B_0^*$ and $B_1^*$ are:

$$B_0^* = \left[\begin{array}{c} 001110001110 \\ 001110001110 \\ \hdashline 001101001101 \\ 001101001101 \\ \hdashline 001011001011 \\ 001011001011 \\ \hdashline 000111000111 \\ 000111000111 \end{array}\right] \begin{matrix} \}\to 1 \\ \\ \}\to 2 \\ \\ \}\to 3 \\ \\ \}\to 4 \\ \end{matrix} \qquad B_1^* = \left[\begin{array}{c} 100011100011 \\ 100011100011 \\ \hdashline 010011010011 \\ 010011010011 \\ \hdashline 001011001011 \\ 001011001011 \\ \hdashline 000111000111 \\ 000111000111 \end{array}\right] \begin{matrix} \}\to 1 \\ \\ \}\to 2 \\ \\ \}\to 3 \\ \\ \}\to 4 \\ \end{matrix}$$

Suppose shares 1, 2 and 3 are used to reveal the secret. Let $x_1$ and $x_2$ be the horizontal displacements of Share 2 and Share 3 relative to Share 1, respectively. Table 8 gives the average contrast in some cases of ( $x_1, x_2$ ). The detailed algorithm of computing average contrast is given in Appendix C.

**Table 8**. Average contrasts with different horizontal shifts

| ($x_1$, $x_2$) | (0, 0) | (1, 0) | (0, 6) | (1, 2) | (6, 1) |
|---|---|---|---|---|---|
| Average contrast | 0.1667 | -0.0306 | 0.0833 | 0.0076 | -0.0167 |

The proof of the security of this construction is relatively straightforward. The contrast issue is more complex since more cases are involved when $k$ shares can be misaligned from each other by various amounts. We leave the scheme as an open problem here.

## 5. Discussions and comparison

First we will compare our STVSS scheme with Yang's scheme in Ref.[9]. By using big and small blocks, Yang et al. proposed a general misalignment tolerant $(k, n)$-VSS scheme, the secret image can be seen when there is a shift. We do an experiment to share the original image in Fig.3 using Yang' scheme and our proposed STVSS scheme when $N_x = 2$ and $N_y = 2$.

**Experiment**: (2, 2) visual secret sharing schemes by using Yang' schemes and our proposed STVSS scheme. For Yang's schemes, we choose the regular masks, i.e. $M_{reg}$ , to arrange the big and small blocks. The percentage of big block is represented by $p_B$ and small block is represented by $p_S$ . Three different percentages: (1) $p_S = 100\%$ and $p_B = 0\%$ , (2) $p_S = p_B = 50\%$ , and (3) $p_S = 0\%$ and $p_B = 100\%$ are tested respectively. For our proposed scheme, we choose $N_x = 2$ and $N_y = 2$. $\overline{a}^{(x,y)}$ is the average contrast of the reconstructed secret image by using STVSS scheme. Experimental result is shown in Fig.4, where ($x$, $y$) denotes different pixel deviations between two shares.

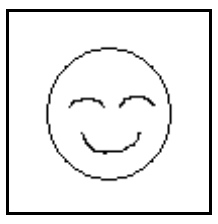

**Fig.** 3. Original image

| $(x, y)$ | Yang[9]( $M_{reg}$) | | | Proposed STVSS | |
|---|---|---|---|---|---|
| | $p_B = 0\%$ $p_S = 100\%$ | $p_B = 50\%$ $p_S = 50\%$ | $p_B = 100\%$ $p_S = 0\%$ | $N_x = 2, N_y = 2$ | $\overline{a}^{(x,y)}$ |
| (0 ,0) | 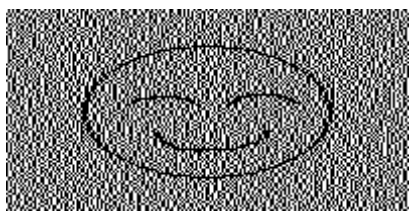 | 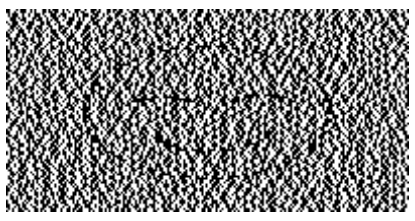 | 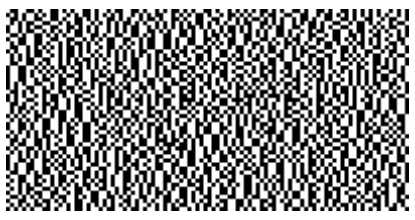 | 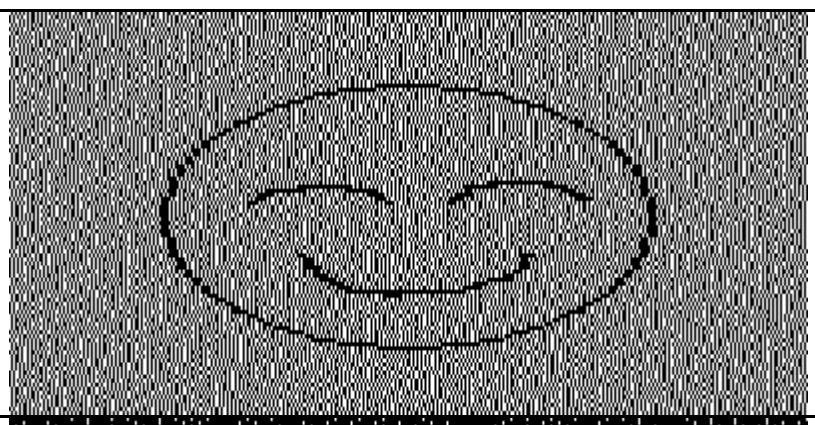 | $\frac{1}{2}$ |
| (1 ,0) | 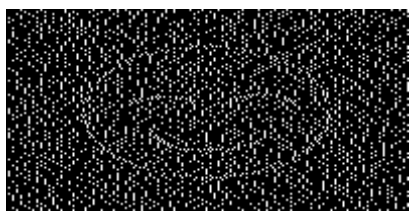 | 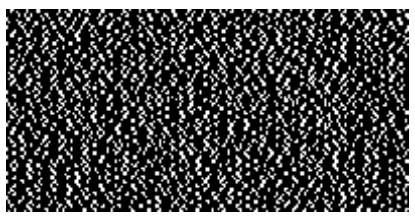 | 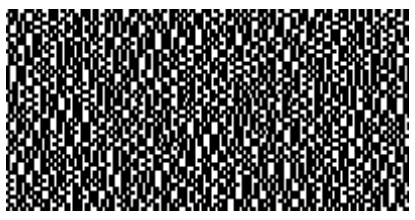 | 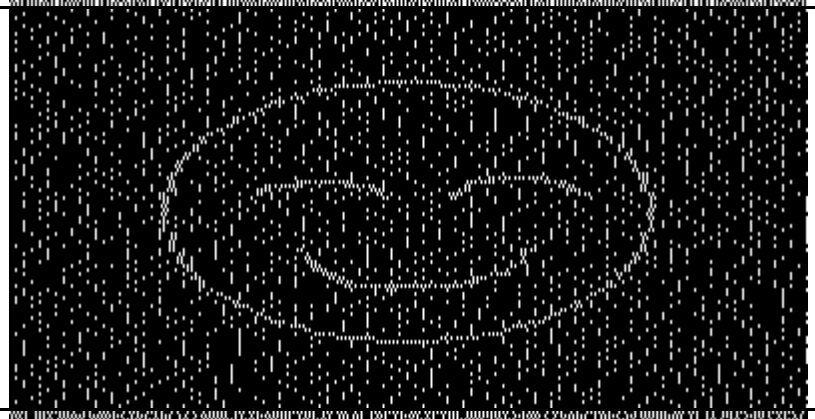 | $-\frac{3}{8}$ |
| (2 ,0) | 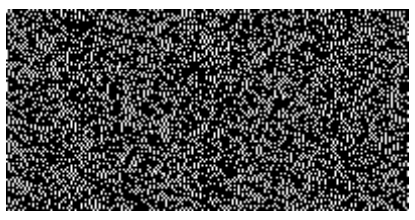 | 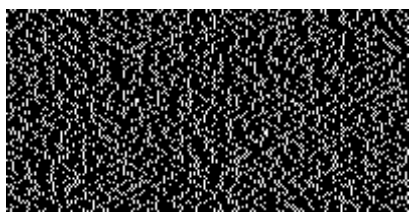 | 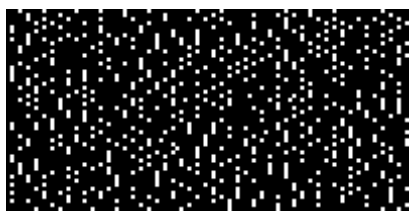 | 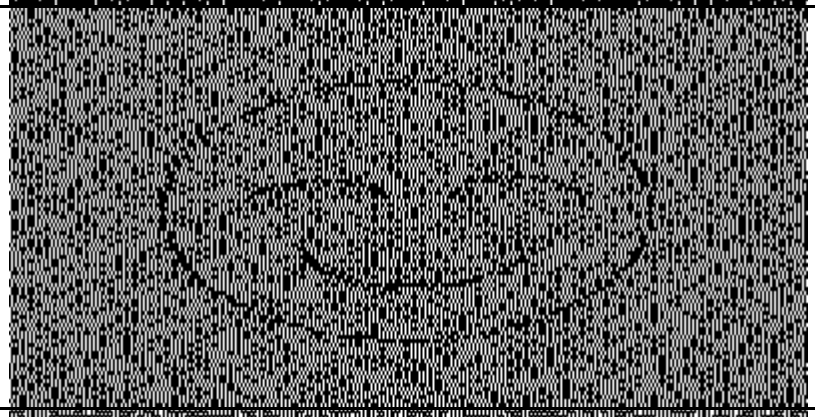 | $\frac{1}{4}$ |
| (0,1) | 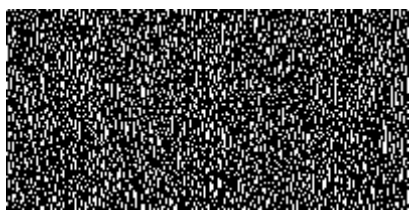 | 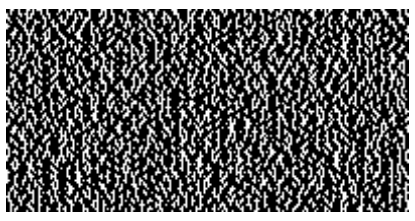 | 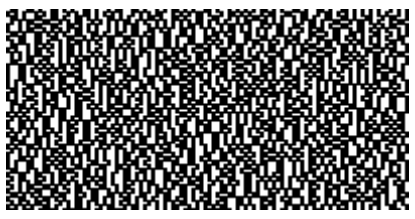 | 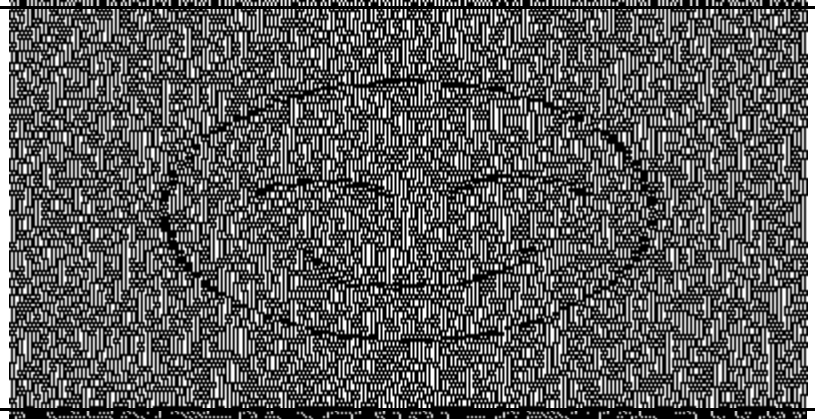 | $\frac{1}{4}$ |
| (1,1) | 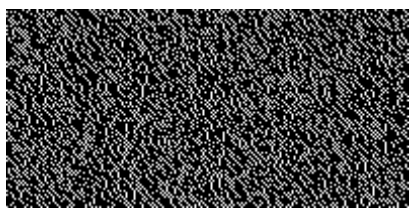 | 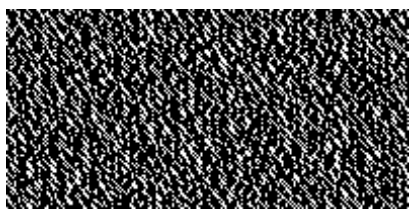 | 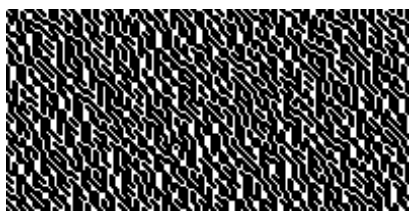 | 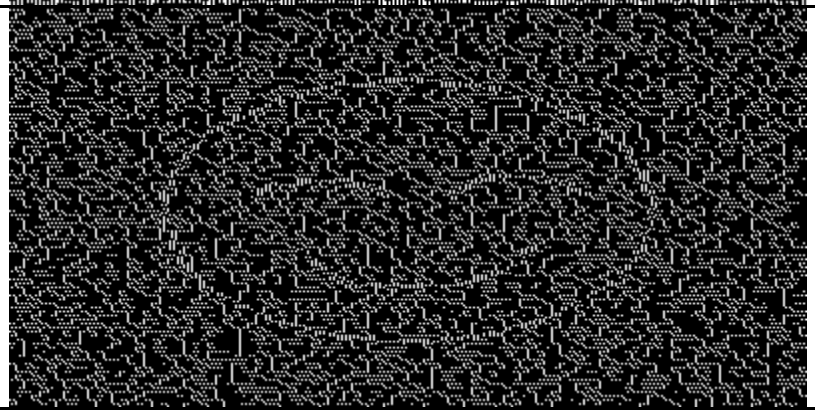 | $-\frac{3}{16}$ |

**Fig.4.** Behaviour of STVSS and the method in [9] when the image contains only thin lines.

Fig.4 demonstrates that the method proposed in Ref. [9] fails to recover the secret image that contains only thin lines. Our proposed scheme can still work in the cases where very little shape redundancy presents in the image. The trade-off is 4 times bigger pixel expansion than Yang's schemes, because $N_x = 2$ and $N_y = 2$. Notice that the proposed STVSS scheme above is compatible with the traditional VSS scheme in terms of the overlay viewing (OR operation), and it can recover the secret image precisely when there is no shift between the shares. When there is a shift, the quality of the recovered image can be evaluated by average contrast, similar to most of the existing VSS schemes. We do not directly compare our method with Yang's in image quality since there is no numerical quality measure given in Ref. [9].

Next we compare the proposed STVSS scheme with the traditional VSS scheme in terms of the contrast and pixel expansion.

***Pixel expansion***

If the pixel expansion of the base VSS scheme is *m*, the pixel expansion of our STVSS scheme would be STVSS scheme is $m^* = N_x N_y m$. Usually, $N_x$ and $N_y$ take values below 4.

***Contrast***

By Corol1ary 2, in traditional (2, *n*)-VSS scheme, with a horizontal shift of *x* pixels ($1 \le x \le m-1$), the average contrast of the recovered image $\overline{a}^{(x,0)} = -\frac{(m-x)(h-l)}{m^2(m-1)} = -\frac{(m-x)}{m(m-1)} a$, here ($1 \le x \le m-1$).

By Theorem 2 and Theorem 3, in (2, *n*)-STVSS scheme. When there is a horizontal shift of *x* ($0 \le x \le m$)pixels and vertical shift $y$ ($0 \le y < N_y$) pixels, the average contrast is

$$\overline{a}^{(0,y)} = \frac{N_y - y}{N_y} \cdot a \ , \ \overline{a}^{(m,y)} = \frac{(N_x - 1)(N_y - y)}{N_x N_y} \cdot a$$

$$\overline{a}^{(x,y)} = -\frac{(N_x m - x)(N_y - y)}{N_x N_y m(m-1)} \cdot a \ , \ 0 < x < m .$$

We resort a conventional $(2,n)$-VSS scheme in [1] to compare with our STVSS scheme.

The basis matrices $B_0$ and $B_1$ are listed as below

$$B_0 = \begin{bmatrix} 100\cdots0 \\ 100\cdots0 \\ \vdots \\ 100\cdots0 \end{bmatrix} , \ B_1 = \begin{bmatrix} 100\cdots0 \\ 010\cdots0 \\ \vdots \\ 000\cdots1 \end{bmatrix}.$$

In (2, $n$)-VSS scheme, the pixel expansion $m = n$, $h = 2, l = 1$. A (2, *n*)-STVSS scheme is generated according to Construction 1. The basis matrices are shown as followings when $N_x = 2$ and $N_y = 2$.

$$B_0^* = \begin{bmatrix} 100\cdots0\,100\cdots0 \\ 100\cdots0\,100\cdots0 \\ 100\cdots0\,100\cdots0 \\ 100\cdots0\,100\cdots0 \\ \vdots \quad \vdots \\ 100\cdots0\,100\cdots0 \\ 100\cdots0\,100\cdots0 \end{bmatrix} , \ B_1^* = \begin{bmatrix} 100\cdots0\,100\cdots0 \\ 100\cdots0\,100\cdots0 \\ 010\cdots0\,010\cdots0 \\ 010\cdots0\,010\cdots0 \\ \vdots \quad \vdots \\ 000\cdots1\,000\cdots1 \\ 000\cdots1\,000\cdots1 \end{bmatrix}.$$

Next, we list the partly values of contrast conventional (2, *n*)-VSS scheme and our (2, *n*)-STVSS scheme in horizontal, vertical and diagonal directions, see table 9, 10, and 11, respectively.

**Table 9**. Comparison the contrast of VSS and STVSS scheme when horizontal shift is involved

| $n$ | $x$=0 | | $x$=1 | | $x$=2 | | $x$=3 | |
|---|---|---|---|---|---|---|---|---|
| | **VSS scheme** | **STVSS scheme** | **VSS scheme** | **STVSS scheme** | **VSS scheme** | **STVSS scheme** | **VSS scheme** | **STVSS scheme** |
| 2 | $\frac{1}{2}$ | $\frac{1}{2}$ | $-\frac{1}{4}$ | $-\frac{3}{8}$ | — | $\frac{1}{4}$ | — | $-\frac{1}{8}$ |
| 3 | $\frac{1}{3}$ | $\frac{1}{3}$ | $-\frac{1}{9}$ | $-\frac{5}{36}$ | $-\frac{1}{18}$ | $-\frac{1}{9}$ | — | $\frac{1}{6}$ |
| 4 | $\frac{1}{4}$ | $\frac{1}{4}$ | $-\frac{1}{16}$ | $-\frac{7}{96}$ | $-\frac{1}{24}$ | $-\frac{1}{16}$ | $-\frac{1}{48}$ | $-\frac{5}{96}$ |

**Table 10**. Comparison the contrast of VSS and STVSS scheme when vertical shift is involved

| $n$ | $y$=0 | | $y$=1 | | $y$=2 | | $y$=3 | |
|---|---|---|---|---|---|---|---|---|
| | **VSS scheme** | **STVSS scheme** | **VSS scheme** | **STVSS scheme** | **VSS scheme** | **STVSS scheme** | **VSS scheme** | **STVSS scheme** |
| 2 | $\frac{1}{2}$ | $\frac{1}{2}$ | — | $\frac{1}{4}$ | — | — | — | — |
| 3 | $\frac{1}{3}$ | $\frac{1}{3}$ | — | $\frac{1}{6}$ | — | — | — | — |
| 4 | $\frac{1}{4}$ | $\frac{1}{4}$ | — | $\frac{1}{8}$ | — | — | — | — |

**Table 11**. Comparison the contrast of VSS and STVSS scheme when diagonal shift is involved

| $n$ | $x$=1,$y$=1, | | $x$=1,$y$=2 | | $x$=2,$y$=1 | | $x$=2,$y$=2 | |
|---|---|---|---|---|---|---|---|---|
| | **VSS scheme** | **STVSS scheme** | **VSS scheme** | **STVSS scheme** | **VSS scheme** | **STVSS scheme** | **VSS scheme** | **STVSS scheme** |
| 2 | $-\frac{1}{4}$ | $-\frac{3}{16}$ | — | — | — | $\frac{1}{8}$ | — | — |
| 3 | $-\frac{1}{18}$ | $-\frac{5}{72}$ | — | — | $-\frac{1}{36}$ | $-\frac{1}{18}$ | — | — |
| 4 | $-\frac{1}{32}$ | $-\frac{7}{192}$ | — | — | $-\frac{1}{48}$ | $-\frac{1}{32}$ | — | — |

For easy lookup and comparison, the traditional $(2, n)$-VSS scheme can still recover the secret image visual when the there is a minor misalignment in the horizontal direction, but completely failed to reveal the secret when vertical or diagonal shifts are involved. Our proposed STVSS scheme still manages to reconstruct the secret image to a degree that visual recognition is possible.

## 6. Conclusion

We have proposed a (2, $n$)-VSS scheme that can tolerate certain displacements between the share transparencies when visually reconstructing the secret image. The quality of the reconstructed image is evaluated by visual inspection and by the average contrast. It is compatible with the traditional VSS when the shares are precisely aligned. When there is a shift, its performance is significantly better than the traditional VSS scheme, and better than duplicating pixels (equivalent to simply enlarging the share images). The trade-off is a larger pixel expansion. Our scheme can reveal the secret to a certain degree even when the original image has very little shape redundancy, such as containing only thin lines. In some cases, the recovered image appears to be a reverse of the original. That is, the black pixels appear white and white pixels appear black. Although the characters or lines in the image are still visible, some "shape information" could be lost, such as the objects in image being hollow or solid. In some cases, this inversion might not be acceptable. However, this phenomenon happens to all traditional and newly proposed schemes, including our own. We are currently investigating this issue and searching for a way towards the "true" shift-tolerance.

## References

[1] M. Naor, A. Shamir. Visual cryptography. Advances in Cryptology-EUROCRYPTO'94, Springer-Verlag, 1995, LNCS, vol. 950, pp.1-12.

[2] K. Kobara, H. Imai. Limiting the visible space visual secret sharing schemes and their application to human identification. ASIACRYPT'96, Springer-Verlag, 1996, LNCS, vol. 1163, pp.185-195.

[3] M.Nakajima,Y.Yamaguchi,Enhancing registration tolerance of extended visual cryptography for natural image, Journal of Electronic imaging, 2004,vol..13,654-622.

[4] Liu, F; Wu, CK; Lin, XJ. The alignment problem of visual cryptography schemes. Designs, Codes and Cryptography, 2009, vol. 50(2), pp. 215-227.

[5] E. R. Verheul, H. C. A. Van Tilborg, Constructions and properties of *k* out of *n* visual secret sharing schemes, Designs, Codes and Cryptography 11 (1997) 179-196.

[6] C. Blundo, A. De Bonis and A. De Santis, "Improved schemes for visual cryptography," *Designs, Codes and Cryptograph*, vol. 24, pp. 255 -278, 2001

[7] Tzong-Lin Wu. Two new visual cryptography schemes: visual multi-secrets sharing scheme and colored visual secret sharing scheme. Taiwan: National Dong Hua University, 2001.

[8] C. Blundo, A. De Santis, Visual cryptography schemes with perfect reconstruction of black pixels, Computers & Graphics, 1998, vol. 22 (4), pp. 449-455.

[9] C. N. Yang, A.G. Peng, T.S. Chen, MTVSS: misalignment tolerant visual secret sharing on resolving alignment difficulty, Signal Processing, 2009, vol.89,pp.1602-1624.

# Appendix A. Notation

| | | |
|---|---|---|
| VSS scheme | $B_0$ and $B_1$ | $n \times m$ basis matrices |
| | $m$ | Pixel expansion |
| | $h$ | Hamming weight of the stacking result of any $k$ out of $n$ rows from matrix in $C_1$ |
| | $l$ | Hamming weight of the stacking result of any $k$ out of $n$ rows from matrix in $C_0$ |
| | $a$ | Relative difference (contrast), $a = (h - l)/m$ |
| STVSS scheme | $x$ | Pixels deviation in vertical |
| | $y$ | Pixels deviation in horizontal |
| | $(x, y)$ | Pixels deviation |
| | $N_x$ | The number of blocks concatenated in the horizontal |
| | $N_y$ | The number of blocks concatenated in the vertical |
| | $B_0^*$ and $B_1^*$ | $N_y n \times N_x m$ basis matrices |
| | $m^*$ | Pixel expansion |
| | $\overline{h}$ | Average Hamming weight of the stacking result of any $k$ out of $n$ rows from matrix in $C_1$ when there is pixel deviation |
| | $\overline{l}$ | Average Hamming weight of the stacking result of any $k$ out of $n$ rows from matrix in $C_0$ when there is pixel deviation |
| | $a^{(0,0)}$ | Contrast when stacked shares without any displacement, $a^{(0,0)} = a = (h - l)/m$ |
| | $\overline{a}^{(x,y)}$ | Average contrast when there is $x$ pixels and $y$ pixels deviation in horizontal and vertical, $\overline{a}^{(x,y)} = -\frac{(N_x m - x)(N_y - y)}{N_x N_y m(m-1)} \cdot a$ |

# Appendix B. Analysis of duplicating pixels and duplicating vectors

Table 1 lists the average contrasts for (2, 2)-VSS schemes under different shift situations. The row "T11" is for the traditional (2, 2)-VSS scheme with no pixel or vector duplications. The (2, 2)-VSS with duplicating *pixels* once in the horizontal direction is shown in row "P21". And the row "V12" is for the (2, 2)-VSS with duplicating *vectors* in the vertical direction, that is, the proposed (2, 2)-STVSS with $(N_x, N_y)$=(1,2). Here, "P" is for "pixel" and "V" is for "vector".

Each table entry, referred to by the column and row indices, is the average contrast in a specific "case": a particular scheme with a particular shift situation. For example, the entry "V21, (2, 0)" is for the case that (2, 2)-VSS with duplicating vectors in the horizontal direction, and there is a horizontal shift of 2 pixels between the shares. Examples to compute average contrast under cases "P22, (1, 1)" and "V22, (1, 1)" are shown in Table 2 and Table 3.

An experiment is performed to share the original secret image in Fig.1. Fig.2 show some recovered secret images in some of the cases listed in Table 1.

**Table 1.** Average contrast of each (2, 2) scheme with each shift distance

| Scheme | $(N_x, N_y)$ | $(x, y)$ | | | | |
|---|---|---|---|---|---|---|
| | | (0, 0) | (1, 0) | (2, 0) | (0, 1) | (1, 1) |
| T11 | 1, 1 | $\frac{1}{2}$ | $-\frac{1}{4}$ | — | — | — |
| P21 | 2, 1 | $\frac{1}{2}$ | $\frac{1}{8}$ | $-\frac{1}{4}$ | — | — |
| V21 | 2, 1 | $\frac{1}{2}$ | $-\frac{3}{8}$ | $\frac{1}{4}$ | — | — |
| P12 | 1, 2 | $\frac{1}{2}$ | $-\frac{1}{4}$ | — | $\frac{1}{4}$ | — |
| V12 | 1, 2 | $\frac{1}{2}$ | $-\frac{1}{4}$ | — | $\frac{1}{4}$ | — |
| P22 | 2, 2 | $\frac{1}{2}$ | $\frac{1}{8}$ | $-\frac{1}{4}$ | $\frac{1}{4}$ | $\frac{1}{16}$ |
| V22 | 2, 2 | $\frac{1}{2}$ | $-\frac{3}{8}$ | $\frac{1}{4}$ | $\frac{1}{4}$ | $-\frac{3}{16}$ |

**Table 2.** Average contrast computing under case “ P22-(1,1)”

| Pixel | Prob. | Share1+Share2 | Share1+ Shifted share2 | Average Hamming weight of the stacked vector | Average contrast |
|---|---|---|---|---|---|
| | 0.5 | $\begin{bmatrix}1100\\1100\end{bmatrix}+\begin{bmatrix}1100\\1100\end{bmatrix}$ | $\begin{bmatrix}1100\\1100\end{bmatrix}+\left[\begin{array}{c\|cccc}1&1&0&0&*\\1&1&0&0&c_1*&c_2&c_3&c_4&c_5\end{array}\right]$ | $\bar{l}=0.5\times(2+w_1)+0.5\times(2+w_2)$ $=2+0.5\times(w_1+w_2)$ | $\bar{a}=\frac{\bar{h}-\bar{l}}{m}=\frac{2.5-2}{8}=\frac{1}{16}$ |
| | 0.5 | $\begin{bmatrix}0011\\0011\end{bmatrix}+\begin{bmatrix}0011\\0011\end{bmatrix}$ | $\begin{bmatrix}0011\\0011\end{bmatrix}+\left[\begin{array}{c\|cccc}0&0&1&1&*\\0&0&1&1&c_1*&c_2&c_3&c_4&c_5\end{array}\right]$ | | |
| | 0.5 | $\begin{bmatrix}1100\\1100\end{bmatrix}+\begin{bmatrix}0011\\0011\end{bmatrix}$ | $\begin{bmatrix}1100\\1100\end{bmatrix}+\left[\begin{array}{c\|cccc}0&0&1&1&*\\0&0&1&1&c_1*&c_2&c_3&c_4&c_5\end{array}\right]$ | $\bar{h}=0.5\times(3+w_1)+0.5\times(2+w_2)$ $=2.5+0.5\times(w_1+w_2)$ | |
| | 0.5 | $\begin{bmatrix}0011\\0011\end{bmatrix}+\begin{bmatrix}1100\\1100\end{bmatrix}$ | $\begin{bmatrix}0011\\0011\end{bmatrix}+\left[\begin{array}{c\|cccc}1&1&0&0&*\\1&1&0&0&c_1*&c_2&c_3&c_4&c_5\end{array}\right]$ | | |

Since $c_1 \ldots c_5$ may be many different kinds of values, we can’t list them all in this table. We show how we compute the average Hamming weight. Consider the first row for white pixel.

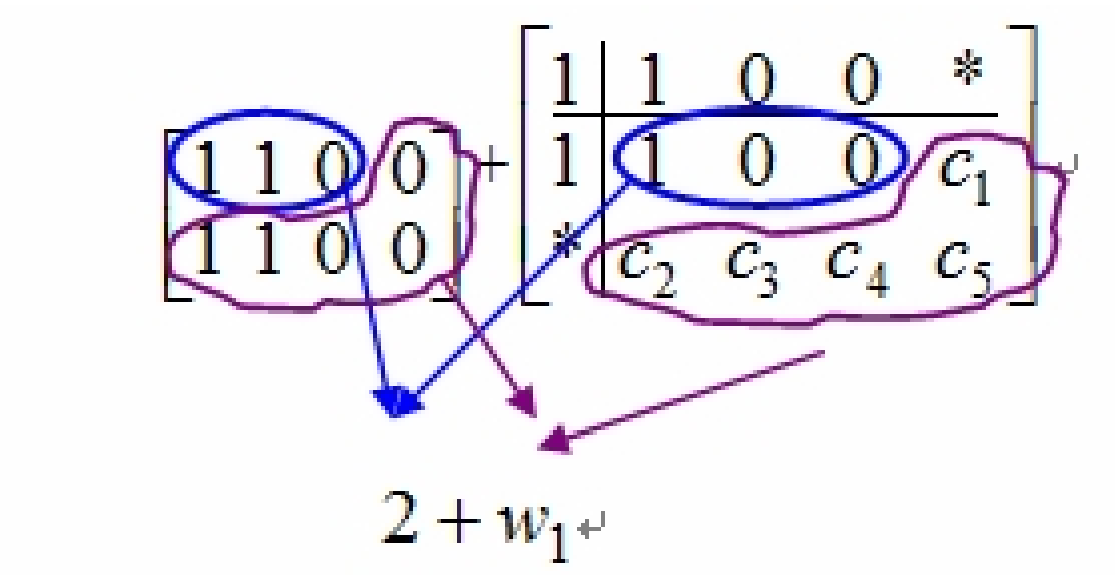

That is the Hamming weight of “110”+ “100” is 2 and we named the average Hamming weight of “01100”+ “$c_1c_2c_3c_4c_5$” is $w_1$. While the average Hamming weight of “10011”+ “$c_1c_2c_3c_4c_5$” is name as $w_2$. We found that $w_1$ and $w_2$ can be reduced to compute the average contrast.

**Table3.** Average contrast computing under case “ V22-(1,1)”

| Pixel | Prob. | Share1+Share2 | Share1+ Shifted share2 | Average Hamming weight of the stacked vector | Average contrast |
|---|---|---|---|---|---|
| | 0.5 | $\begin{bmatrix}1010\\1010\end{bmatrix}+\begin{bmatrix}1010\\1010\end{bmatrix}$ | $\begin{bmatrix}1010\\1010\end{bmatrix}+\left[\begin{array}{c\|cccc}1&0&1&0&*\\1&0&1&0&c_1*&c_2&c_3&c_4&c_5\end{array}\right]$ | $\bar{l}=0.5\times(3+w_1)+0.5\times(3+w_2)$ $=3+0.5\times(w_1+w_2)$ | $\bar{a}=\frac{\bar{h}-\bar{l}}{m}=\frac{1.5-3}{8}=-\frac{3}{16}$ |
| | 0.5 | $\begin{bmatrix}0101\\0101\end{bmatrix}+\begin{bmatrix}0101\\0101\end{bmatrix}$ | $\begin{bmatrix}0101\\0101\end{bmatrix}+\left[\begin{array}{c\|cccc}0&1&0&1&*\\0&1&0&1&c_1*&c_2&c_3&c_4&c_5\end{array}\right]$ | | |
| | 0.5 | $\begin{bmatrix}1010\\1010\end{bmatrix}+\begin{bmatrix}0101\\0101\end{bmatrix}$ | $\begin{bmatrix}1010\\1010\end{bmatrix}+\left[\begin{array}{c\|cccc}0&1&0&1&*\\0&1&0&1&c_1*&c_2&c_3&c_4&c_5\end{array}\right]$ | $\bar{h}=0.5\times(2+w_1)+0.5\times(1+w_2)$ $=1.5+0.5\times(w_1+w_2)$ | |
| | 0.5 | $\begin{bmatrix}0101\\0101\end{bmatrix}+\begin{bmatrix}1010\\1010\end{bmatrix}$ | $\begin{bmatrix}0101\\0101\end{bmatrix}+\left[\begin{array}{c\|cccc}1&0&1&0&*\\1&0&1&0&c_1*&c_2&c_3&c_4&c_5\end{array}\right]$ | | |

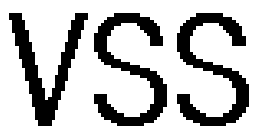

**Fig.1** Original Secret image

| $(x, y)$ | Duplicating vectors | Duplicating pixels |
|---|---|---|
| | $B_0 = \begin{bmatrix} 1 & 0 \\ 1 & 0 \end{bmatrix} \Rightarrow B_0 = \begin{bmatrix} 1 & 0 & 1 & 0 \\ 1 & 0 & 1 & 0 \\ \hdashline 1 & 0 & 1 & 0 \\ 1 & 0 & 1 & 0 \end{bmatrix}$ <br> $B_1 = \begin{bmatrix} 1 & 0 \\ 0 & 1 \end{bmatrix} \Rightarrow B_1 = \begin{bmatrix} 1 & 0 & 1 & 0 \\ 1 & 0 & 1 & 0 \\ \hdashline 0 & 1 & 0 & 1 \\ 0 & 1 & 0 & 1 \end{bmatrix}$ | $B_0 = \begin{bmatrix} 1 & 0 \\ 1 & 0 \end{bmatrix} \Rightarrow B_0 = \begin{bmatrix} 1 & 1 & 0 & 0 \\ 1 & 1 & 0 & 0 \\ \hdashline 1 & 1 & 0 & 0 \\ 1 & 1 & 0 & 0 \end{bmatrix}$ <br> $B_1 = \begin{bmatrix} 1 & 0 \\ 0 & 1 \end{bmatrix} \Rightarrow B_1 = \begin{bmatrix} 1 & 1 & 0 & 0 \\ 1 & 1 & 0 & 0 \\ \hdashline 0 & 0 & 1 & 1 \\ 0 & 0 & 1 & 1 \end{bmatrix}$ |
| (0, 0) | 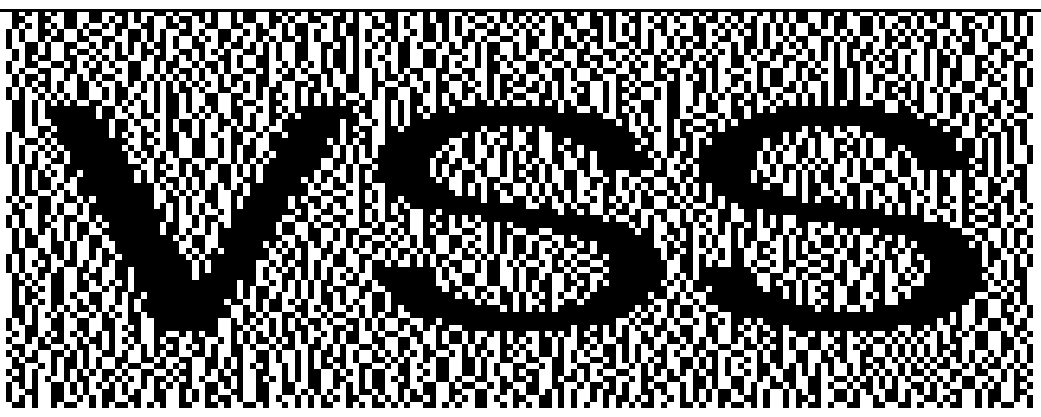 | 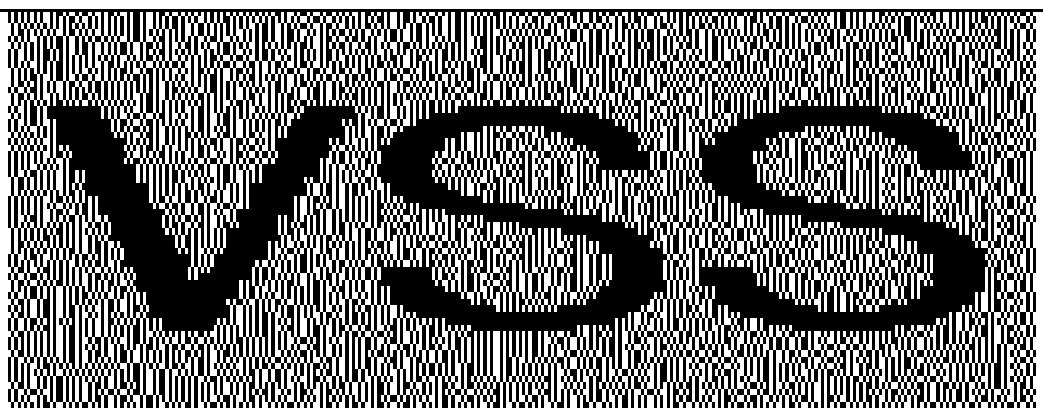 |
| (1, 0) | 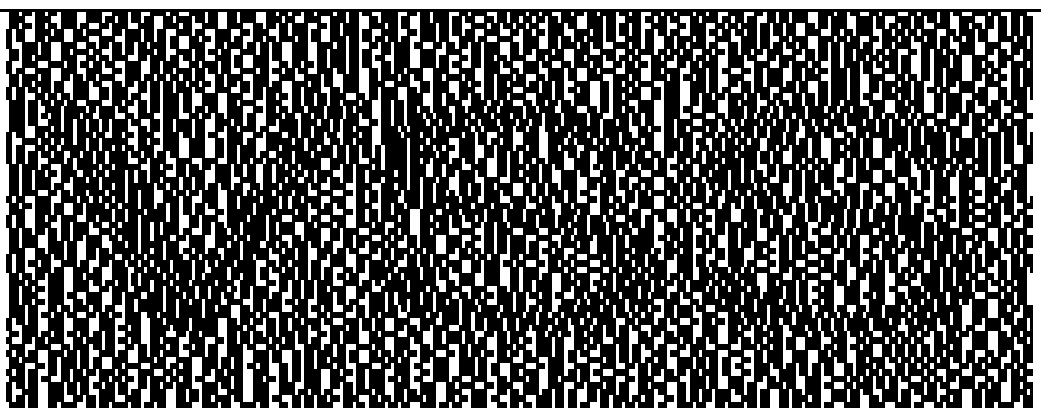 | 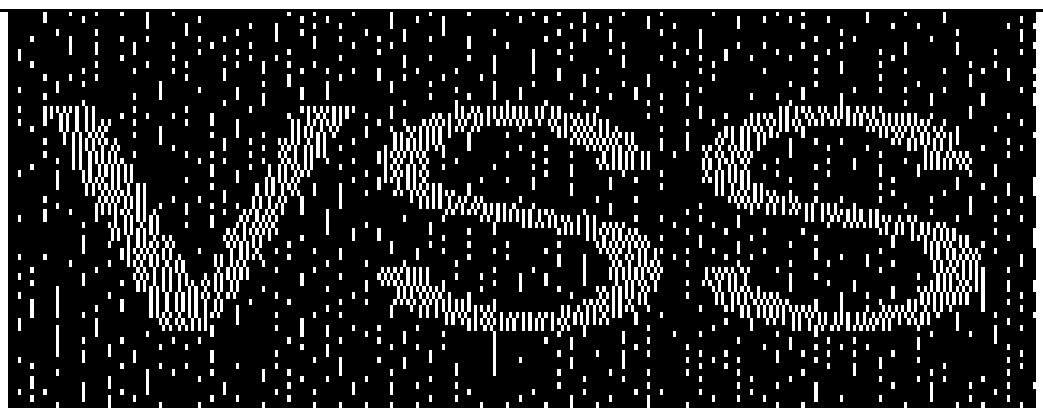 |
| (2, 0) | 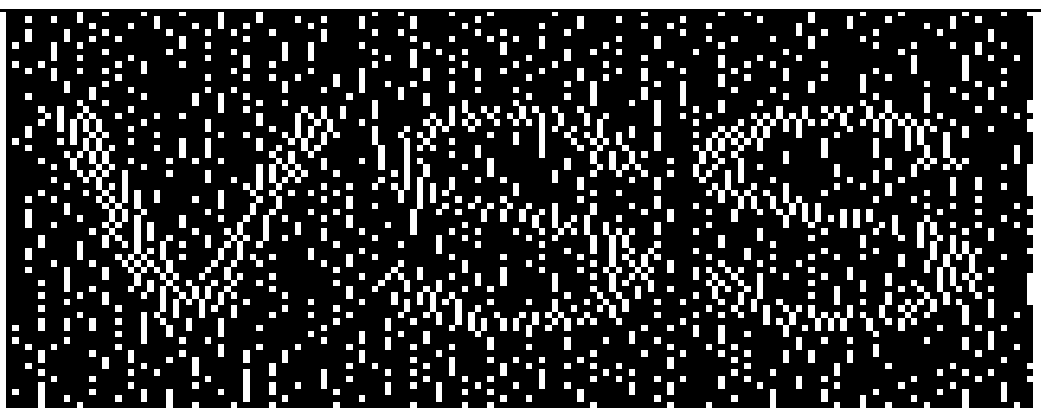 | 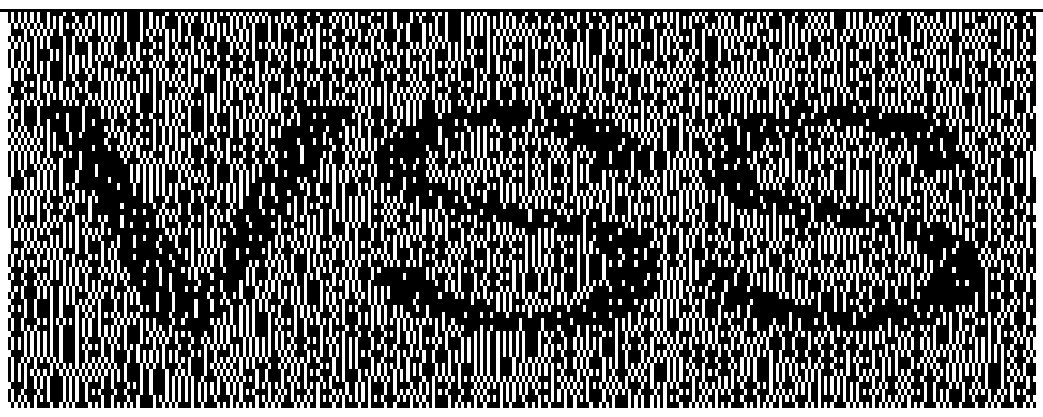 |
| (0, 1) | 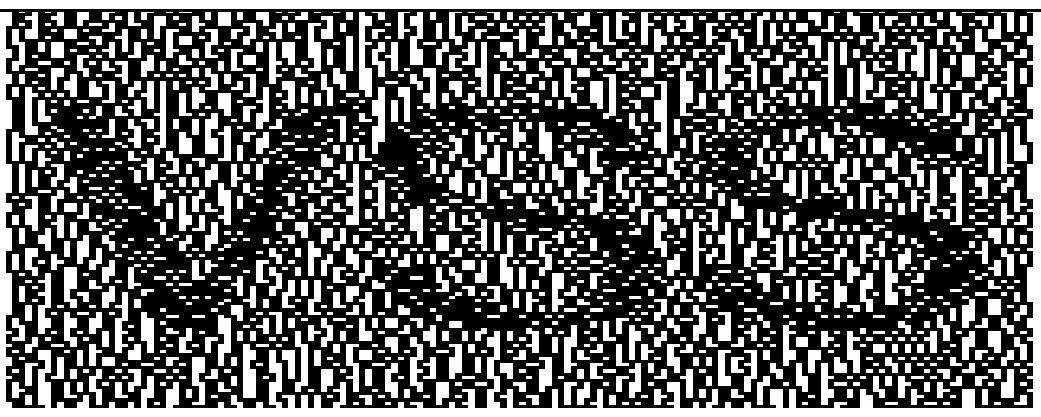 | 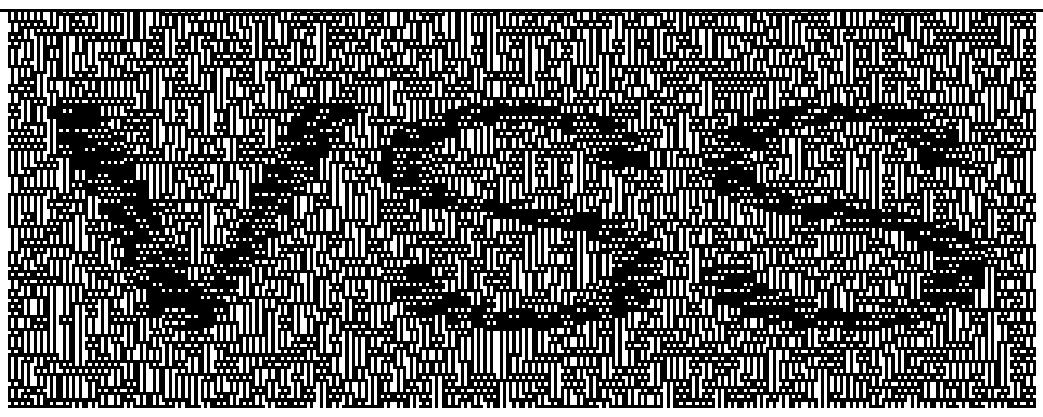 |
| (1, 1) | 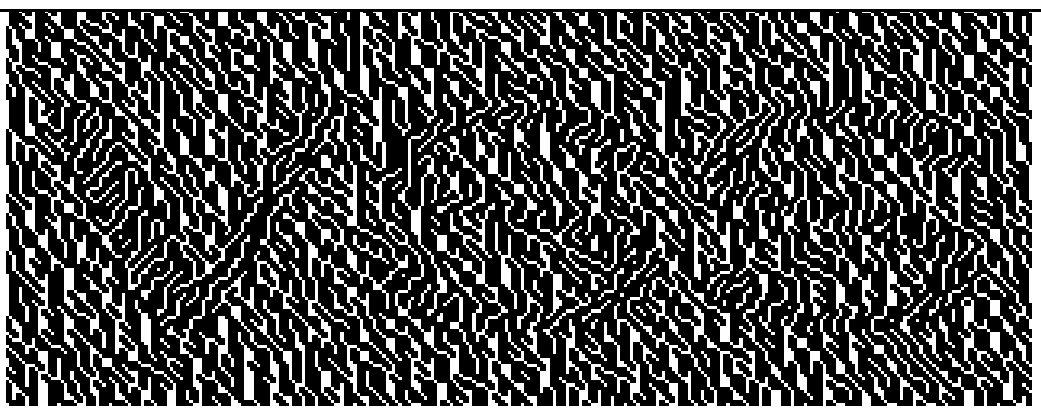 | 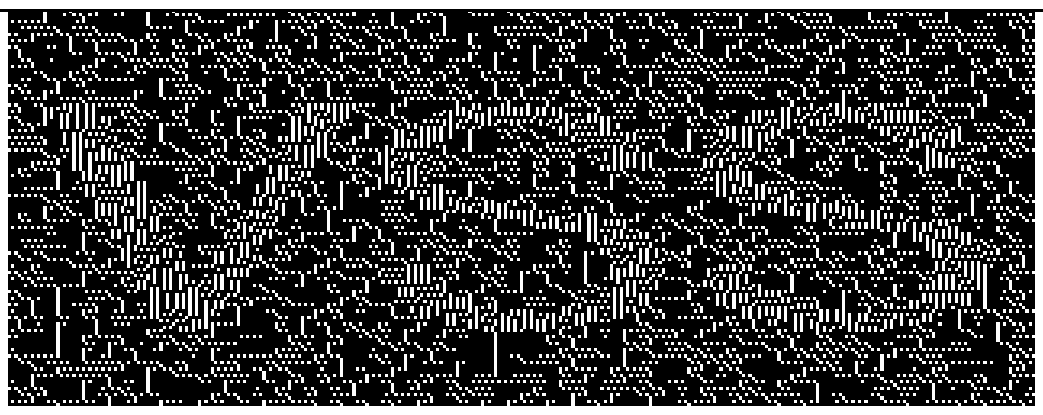 |

**Fig.2** Comparison of the schemes with one example secret image. $(x, y)$ means share 2 is shifted to left by $x$ pixels and down by $y$ pixels according to share 1.

## Appendix C. Three kinds of permutation principles and algorithm to compute the average contrast

Using three different permutation methods, we shall give contrast comparison of a (2, 3)-STVSS scheme in the horizontal direction. The matrices $B_0^*$ and $B_1^*$ of a horizontal shift tolerant (2, 3)-STVSS scheme are listed bellows, also see example 3.

$$B_0^* = \left[\begin{array}{ccc|ccc} 0 & 1 & 1 & 0 & 1 & 1 \\ 0 & 1 & 1 & 0 & 1 & 1 \\ 0 & 1 & 1 & 0 & 1 & 1 \end{array}\right]\begin{array}{l} \to 1 \\ \to 2 \\ \to 3 \end{array} \qquad B_1^* = \left[\begin{array}{ccc|ccc} 0 & 1 & 1 & 0 & 1 & 1 \\ 1 & 1 & 0 & 1 & 1 & 0 \\ 1 & 0 & 1 & 1 & 0 & 1 \end{array}\right]\begin{array}{l} \to 1 \\ \to 2 \\ \to 3 \end{array}$$

Let $C_0^*$ and $C_1^*$ be the collection of all matrices obtained by permuting columns of $B_0^*$ and $B_1^*$ according to following there methods :

*Permutation Method 1* : Let $C_0^*$ and $C_1^*$ be the collection of all matrices obtained by permuting full columns of $B_0^*$ and $B_1^*$. In the situation, there exists 120 columns permutation in basis matrices $C_0^*$ and $C_1^*$,respectively.

*Permutation Method 2* : Let $C_0^*$ and $C_1^*$ be the collection of all matrices obtained by permuting three columns on the left side of marked symbol " | " and the three columns on the right side, respectively. that is , the basis matrices $C_0^*$ and $C_1^*$ include 36 matrices, respectively.

We now list some columns permutation of $C_0^*$ and $C_1^*$ as bellows:

$$C_0^* = \{\left[\begin{array}{ccc|ccc} 0 & 1 & 1 & 0 & 1 & 1 \\ 0 & 1 & 1 & 0 & 1 & 1 \\ 0 & 1 & 1 & 0 & 1 & 1 \end{array}\right], \left[\begin{array}{ccc|ccc} 1 & 0 & 1 & 0 & 1 & 1 \\ 1 & 0 & 1 & 0 & 1 & 1 \\ 1 & 0 & 1 & 0 & 1 & 1 \end{array}\right], \ldots, \left[\begin{array}{ccc|ccc} 0 & 1 & 1 & 1 & 1 & 0 \\ 0 & 1 & 1 & 1 & 1 & 0 \\ 0 & 1 & 1 & 1 & 1 & 0 \end{array}\right]\}$$

$$C_1^* = \{\left[\begin{array}{ccc|ccc} 0 & 1 & 1 & 0 & 1 & 1 \\ 1 & 1 & 0 & 1 & 1 & 0 \\ 1 & 0 & 1 & 1 & 0 & 1 \end{array}\right], \left[\begin{array}{ccc|ccc} 1 & 1 & 1 & 0 & 1 & 1 \\ 1 & 0 & 0 & 1 & 1 & 0 \\ 0 & 1 & 1 & 1 & 0 & 1 \end{array}\right], \ldots, \left[\begin{array}{ccc|ccc} 0 & 1 & 1 & 1 & 1 & 0 \\ 1 & 1 & 1 & 0 & 1 & 1 \\ 1 & 0 & 1 & 1 & 0 & 1 \end{array}\right]\}$$

*Permutation Method 3*: Let $C_0^*$ and $C_1^*$ be the collection of all matrices obtained by permuting three columns on the left side of marked symbol " | " and the three columns on the right side with the same sequence of permutation. The basis matrices $C_0^*$ and $C_1^*$ are stipulated 6 columns permutation, respectively.

The 6 columns permutation of $C_0^*$ and $C_1^*$ in the method are listed in example 5.

Now we compute the average contrast in three columns permutation above when there is deviations by using algorithm of computing average contrast, the experiment results are showed in Table 1.

**Table 1**. Comparison of the average contrast in different permutations.

| | $x$=0 | $x$=1 | $x$=2 | $x$=3 |
|---|---|---|---|---|
| Permutation Method 1 | 1/3 | -1/18 | -4/90 | -1/30 |
| Permutation Method 2 | 1/3 | -1/9 | -1/18 | 0 |
| Permutation Method 3 | 1/3 | -5/36 | -1/9 | 1/6 |

From the table 1, it is easily shown that the average contrast of Permutation Method 3 is optimal.

**Algorithm of computing average contrast**

Input: $B_0^*$ and $B_1^*$, Permutation method , Horizontal deviation $x$

Output: Average contrast

Step1: The matrix collections $C_0^*$, $C_1^*$ are generated by applying chosen permutation method on $B_0^*$ and $B_1^*$.

Step2: For each matrix $T_0$, $T_1$ from $C_0^*$, $C_1^*$.

Choose any $k$ rows $wr_{i_1},\ldots wr_{i_k}$ from $T_0$ to recover the white pixel;

Choose any $k$ rows $br_{i_1},\ldots br_{i_k}$ from $T_1$ to recover the black pixel;

Shift $wr_{i_q}$ and $br_{i_q}$ to left by $x$ pixels, there will be $x$ pixels $c_1 \cdots c_x$ shifted in.

For all $c_1 \cdots c_x$, compute Hamming weight and add it to the total number.

$hw_c_1 \cdots c_x$ +=H( $wr_{i_1}$ OR… OR *shifted* $wr_{i_q}$ OR …OR $wr_{i_k}$ ) ;

$hb_c_1 \cdots c_x$ +=H( $br_{i_1}$ OR… OR *shifted* $br_{i_q}$ OR …OR $br_{i_k}$ );

Step3: Compute average contrast $\bar{a} = \dfrac{\sum_{c_1 \cdots c_x} p_{c_1 \cdots c_x}(hb_c_1 \cdots c_x - hw_c_1 \cdots c_x)}{|C_0^*|}$.

Here $p_{c_1 \cdots c_x}$ denote the probability that a string $c_1 \cdots c_x$ is shifted in. $hb_c_1 \cdots c_x$ and $hw_c_1 \cdots c_x$ denote the total Hamming weight when $c_1 \cdots c_x$ are shifted in.

## Appendix D. Contrast a (2, 3)-STVSS scheme in horizontal direction

In example 3, we give a method to construct a horizontal shift tolerant (2, 3)-VSS scheme, the basis matrices $B_0^*$ and $B_1^*$ are

$$B_0^* = \left[\begin{array}{ccc|ccc} 0 & 1 & 1 & 0 & 1 & 1 \\ 0 & 1 & 1 & 0 & 1 & 1 \\ 0 & 1 & 1 & 0 & 1 & 1 \end{array}\right]\begin{array}{l}\to 1\\ \to 2\\ \to 3\end{array} \qquad B_1^* = \left[\begin{array}{ccc|ccc} 0 & 1 & 1 & 0 & 1 & 1 \\ 1 & 1 & 0 & 1 & 1 & 0 \\ 1 & 0 & 1 & 1 & 0 & 1 \end{array}\right]\begin{array}{l}\to 1\\ \to 2\\ \to 3\end{array}$$

We now check the construction of example 3 against the contrast condition.

Without loss of generality, we suppose Share 1 (denoted by $S_1$) and Share 3 (denoted by $S_3$) are stacked to reconstruct the secret image. The original white (resp. black) pixel is reconstructed by stacking $B_0^*[1]$ and $B_0^*[3]$ (resp. $B_1^*[1]$ and $B_1^*[3]$). We assume that $S_3$ is shifted to the left by $x$ pixels with respect to $S_1$. In other words, $x$ pixels of $B_0^{LR}[3]$ (resp. $B_1^*[3]$) of $S_3$ are shifted out and $x$ adjacent pixels of $S_3$ are shifted in to fill the space. Let $c_1 \mathbf{K} c_x$ represent the $x$ pixels shifted in. When x=3, three pixels $c_1c_2c_3 \in \{011, 110, 101\}$ are shifted in. If $c_1c_2c_3$ = 110, $B_0^*[3] = [0 \;\; 1 \;\; 1 \;\; 0 \;\; 1 \;\; 1]$ will become

$$[0 \;\; 1 \;\; 1 \,\vdots\, 0 \;\; 1 \;\; 1 \;\; c_1 \;\; c_2 \;\; c_3] = [0 \;\; 1 \;\; 1 \,\vdots\, 0 \;\; 1 \;\; 1 \;\; 1 \;\; 1 \;\; 0],$$

where the three pixels 011 to the left of the dotted line are shifted out.

Some examples are given in Table 1 and Table 2 to demonstrate the construction process for white and black pixels of the original image. Each table has four columns corresponding to four values of *x*.

**Table 1.** Reconstruction process of a white original pixel with horizontal displacement of *x* pixels

| | *x*=0 | *X*=1<br>$c_1 = 0$ | *x*=2<br>$c_1c_2 = 10$ | *x*=3<br>$c_1c_2c_3 = 110$ |
|---|---|---|---|---|
| $B_0^*[1]$ | $[0\ 1\ 1\ 0\ 1\ 1]$ | $[0\ 1\ 1\ 0\ 1\ 1]$ | $[0\ 1\ 1\ 0\ 1\ 1]$ | $[0\ 1\ 1\ 0\ 1\ 1]$ |
| $B_0^*[3]$ | $[0\ 1\ 1\ 0\ 1\ 1]$ | $[0 \vdots 1\ 1\ 0\ 1\ 1\ 0]$ | $[0\ 1 \vdots 1\ 0\ 1\ 1\ 1\ 0]$ | $[0\ 1\ 1 \vdots 0\ 1\ 1\ 1\ 1\ 0]$ |
| $V^0 = B_0^*[1]$<br>$+ B_0^*[3]$ | $[0\ 1\ 1\ 0\ 1\ 1]$ | $[1\ 1\ 1\ 1\ 1\ 1]$ | $[1\ 1\ 1\ 1\ 1\ 1]$ | $[0\ 1\ 1\ 1\ 1\ 1]$ |
| $H(V^0)$ | 4 | 6 | 6 | 5 |

**Table 2.** Reconstruction process of a black original pixel with horizontal displacement of *x* pixels

| | $x$=0 | $x$=1<br>$c_1 = 0$ | $x$=2<br>$c_1c_2 = 10$ | $x$=3<br>$c_1c_2c_3 = 110$ |
|---|---|---|---|---|
| $B_1^*[1]$ | $[0\ 1\ 1\ 0\ 1\ 1]$ | $[0\ 1\ 1\ 0\ 1\ 1]$ | $[0\ 1\ 1\ 0\ 1\ 1]$ | $[0\ 1\ 1\ 0\ 1\ 1]$ |
| $B_1^*[3]$ | $[1\ 0\ 1\ 1\ 0\ 1]$ | $[1 \vdots 0\ 1\ 1\ 0\ 1\ 0]$ | $[1\ 0 \vdots 1\ 1\ 0\ 1\ 1\ 0]$ | $[1\ 0\ 1 \vdots 1\ 0\ 1\ 1\ 1\ 0]$ |
| $V^1 = B_1^*[1] + B_1^*[3]$ | $[1\ 1\ 1\ 1\ 1\ 1]$ | $[0\ 1\ 1\ 0\ 1\ 1]$ | $[1\ 1\ 1\ 1\ 1\ 1]$ | $[1\ 1\ 1\ 1\ 1\ 1]$ |
| $H(V^1)$ | 6 | 4 | 6 | 6 |

The contrast can be computed from the above tables. For example, when $x$=3, $c_1c_2c_3 = 110$, we have $H(V^0) = 5$ and $H(V^1) = 6$. The contrast is $a = \frac{H(V^1) - H(V^0)}{m^*} = \frac{6-5}{6} = \frac{1}{6}$.

# Appendix F. A shift tolerant (2, 3)-VSS scheme

This is the experimental result of shift tolerant (2, 3)-VSS scheme with $N_x = 2$ and $N_y = 2$. The pixel expansion is $m = 2 \cdot 2 \cdot 3 = 12$ and the sub-pixels are arranged in a $2 \times 6$ pattern. In Fig.1: (a) is the secret image. (b)-(d) are three shares. (e) is the stacking result when share 1 and share 2 are stacked precisely. (f) is the stacking result when share 2 is shifted up by 1 pixel according to share 1. (g) is the stacking result when share 2 is shifted left by 1 pixel according to share 1. (h) is the stacking result when share 2 is shifted left by 2 pixels according to share 1.(i) is the stacking result when share 2 is shifted left by 3 pixels according to share 1. (j) is the stacking result when share 2 is shifted right and down respectively by 1 pixel according to share 1. (k) is the stacking result when share 2 is shifted right and up respectively by 1 pixel according to share 1.

(a) Secret

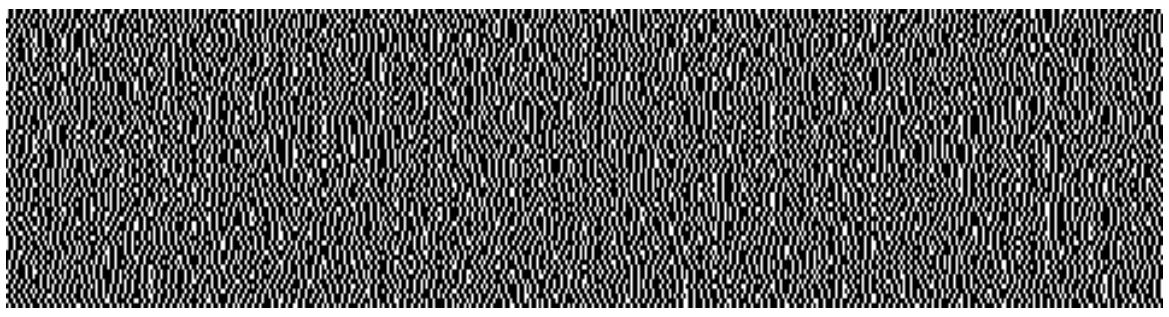

(b) Share 1

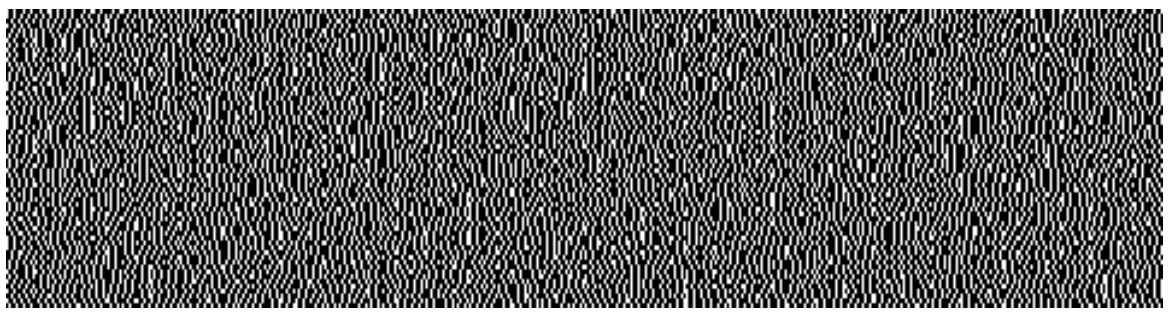

(c) Share 2

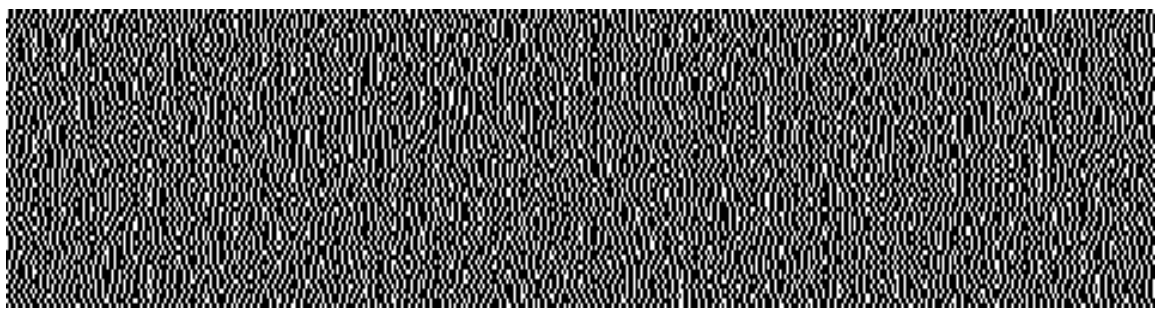

(d) Share 3

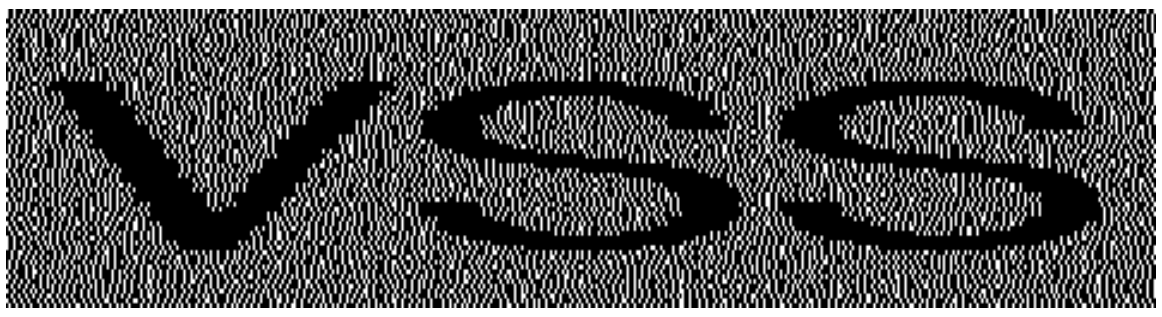

(e) $S_1$ and $S_2$ are stacked precisely

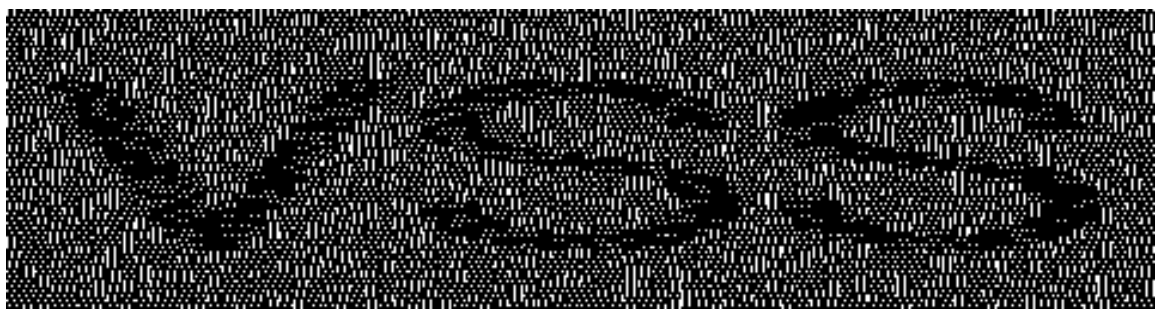

(f) $S_2$ is shifted up by 1 pixel

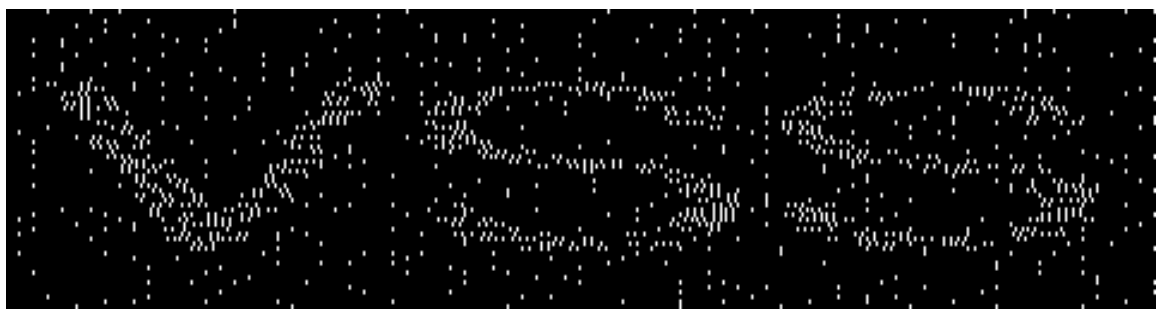

(g) $S_2$ is shifted left by 1 pixel

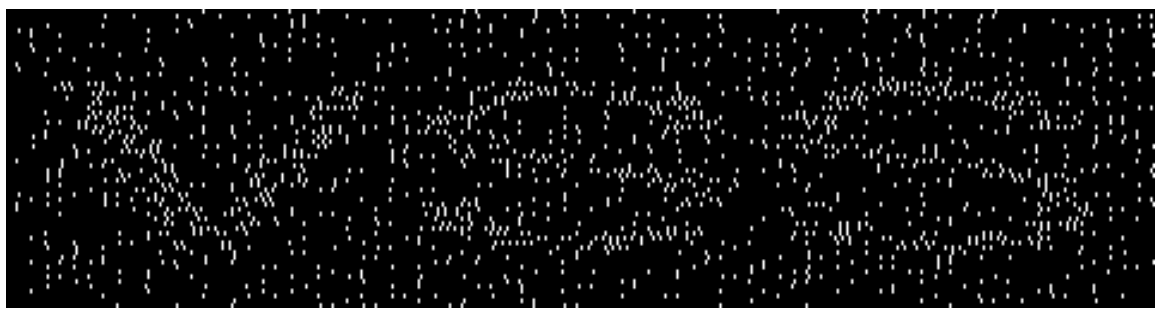

(h) $S_2$ is shifted left by 2 pixels

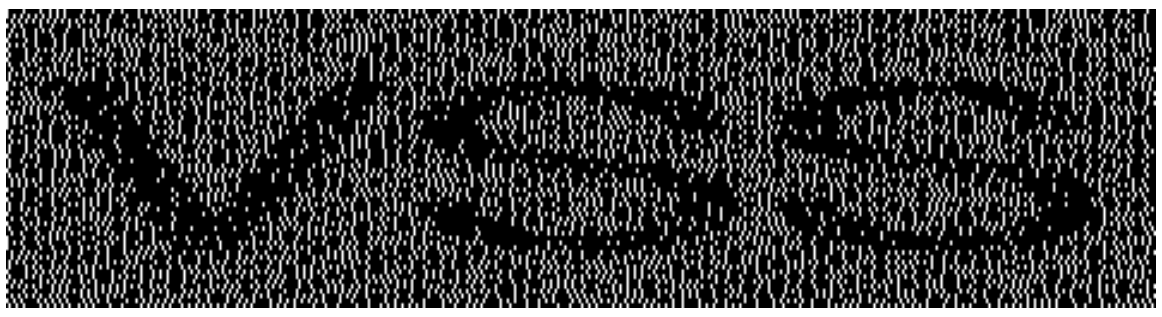

(i) $S_2$ is shifted left by 3 pixels

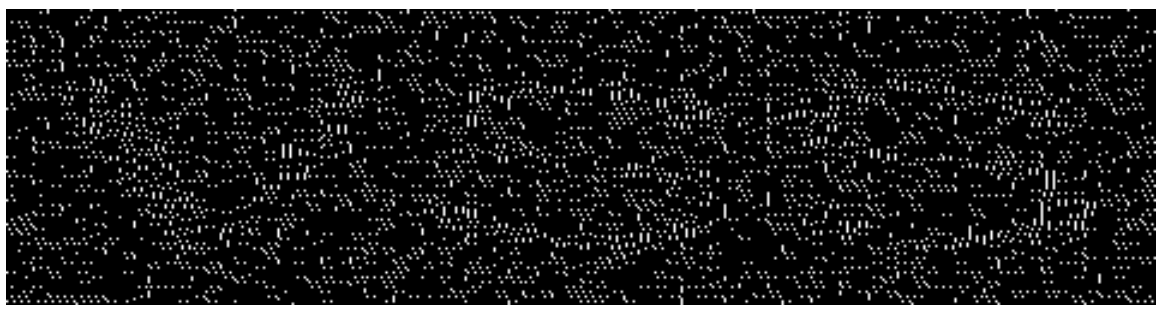

(j) $S_2$ is shifted right and down by 1 pixel

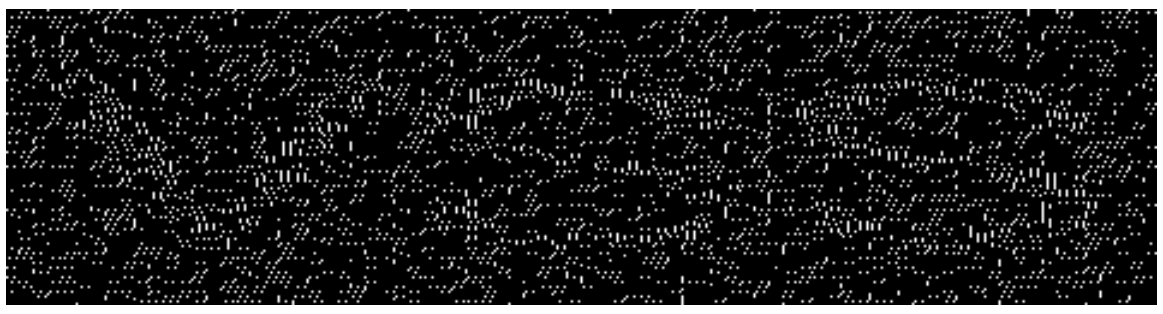

(k) $S_2$ is shifted right and up by 1 pixel

**Fig.1** Experimental result of shift tolerant (2, 3)-VSS scheme with $N_x = 2$ and $N_y = 2$